\documentclass[letterpaper]{JHEP3}
\usepackage{amsmath}
\usepackage{amssymb}
\usepackage{bm}
\usepackage{multirow}
\usepackage{graphicx}
\usepackage{cite}

\numberwithin{equation}{section}

\newcommand{\mb}{\mathbf}
\newcommand{\eg}{\emph{e.g.}}
\newcommand{\ie}{\emph{i.e.}}
\newcommand{\ol}{\overline}

\newcommand{\be}{\begin{equation}}
\newcommand{\ee}{\end{equation}}
\newcommand{\ben}{\begin{equation*}}
\newcommand{\een}{\end{equation*}}
\newcommand{\bea}{\begin{eqnarray}}
\newcommand{\eea}{\end{eqnarray}}
\newcommand{\bean}{\begin{eqnarray*}}
\newcommand{\eean}{\end{eqnarray*}}
\newcommand{\nno}{\nonumber}
\newcommand{\imp}{\Rightarrow}
\newcommand{\goesto}{\rightarrow}

\newcommand{\bTi}{\begin{itemize} \setlength{\itemsep}{-.1cm}}
\newcommand{\eTi}{\end{itemize}}

\newcommand{\pd}{\partial}

\renewcommand{\Im}{{\mbox{Im}}}
\renewcommand{\Re}{{\mbox{Re}}}

\newcommand{\DFYZ}{{D\cF \hspace{-.1cm}\cdot\hspace{-.1cm}\bar{Y} \hspace{-.14cm}\cdot\hspace{-.11cm} \bar{Z} }}
\newcommand{\bDFYZ}{{ \ol{D\cF} \hspace{-.1cm}\cdot\hspace{-.1cm}{Y} \hspace{-.14cm}\cdot\hspace{-.11cm} {Z} }}
\newcommand{\ccdot}{\hspace{-.1cm}\cdot\hspace{-.1cm}}

\newcommand{\bse}{\begin{subequations}}
\newcommand{\ese}{\end{subequations}}
\newcommand{\XYZ}{{X,Y_{\hat{i}},Z_{\hat{a}}}}
\newcommand{\DFab}{{D_{\hat{a}}\cF_{\hat{b}\hat{c}\hat{d}}}}
\newcommand{\Fab}{{{\cal F}_{\hat{a}\hat{b}\hat{c}}}}
\newcommand{\cF}{{\cal F}}
\newcommand{\tW}{{\tilde{W}}}

\newcommand{\mbc}{{\mathbb{C}}}
\newcommand{\mbr}{{\mathbb{R}}}
\newcommand{\mbz}{{\mathbb{Z}}}
\newcommand{\Y}[1]{{Y_{\hat{#1}}}}
\newcommand{\Z}[1]{{Z_{\hat{#1}}}}
\newcommand{\D}[1]{{D_{\hat{#1}}}}
\newcommand{\cY}[1]{{\bar{Y}_{\hat{\bar{#1}}}}}
\newcommand{\cZ}[1]{{\bar{Z}_{\hat{\bar{#1}}}}}
\newcommand{\cX}{{\bar{X}}}

\newcommand{\ds}{\displaystyle}

\title{Type IIB Flux Vacua at Large Complex Structure}

\author{Tudor Dan Dimofte \\ California Institute of Technology, Pasadena, CA 91125, USA \\ E-mail: \email{tdd@theory.caltech.edu}}

\abstract{We study models of stabilization near large complex structure in type IIB O3/O7 flux compactifications. We consider a special family of examples with a single nonvanishing Yukawa coupling in the large-complex-structure limit, which allows us to study all possible stable vacua of the tree-level no-scale potential very explicitly. We find that, by tuning fluxes, both supersymmetric and nonsupersymmetric vacua can be realized at almost any point in the large-complex-structure moduli space of one-, two-, and three-parameter models. We also consider the effects of stringy corrections on tree-level vacua. We argue quite generally that, in certain regimes, both supersymmetric and nonsupersymmetric tree-level vacua could serve as consistent, controllable foundations for full stabilization beyond tree level (including K\"ahler moduli), leading to either AdS or dS cosmological constants. We show how to achieve these regimes in our models. Finally, we discuss some implications of minimizing at tree level the no-scale form of the scalar potential, versus other potentials used in statistical studies.}

\preprint{CALT-68-2688 \\ IPMU-08-0033}

\begin{document}

\section{Introduction}

Type IIB flux compactifications, particularly the ${\cal N}=1$ O3/O7 orientifold compactifications developed in \cite{GKP} and reviewed most recently in \cite{Denef}, have become increasingly popular in recent years. Part of their appeal stems from their ability to stabilize all geometric moduli at large volume \cite{KKLT, Denef:2004dm, Denef:2005mm, BBCQ, BHP, CQS}, as well as their potential for realizing stringy inflation \cite{KKLMMT, BlancoPillado:2004ns, BDKM, CQ-infl, KP}. Inspired by the seminal work of KKLT \cite{KKLT}, many IIB stabilization scenarios assume that the complex-structure moduli and the axion-dilaton are fixed supersymmetrically at tree level using fluxes,%
\footnote{Throughout this paper we shall use ``tree level'' to mean both string tree level and leading order in the $\alpha'$ (\ie\ large volume) expansion.} %
and then proceed to study the more intricate problem of K\"ahler (or open string) modulus stabilization using stringy corrections. In this paper, we momentarily shift the focus back to tree level. We describe a special class of tree-level models that can be stabilized near large complex structure both supersymmetrically and non-supersymmetrically, and evaluate their practicability as a basis for more complete stabilization scenarios.

The models we present have $n=1,\,2,$ and $3$ complex-structure parameters. We require that they have a single nonvanishing Yukawa coupling in the large-complex-structure limit. Although this requirement is somewhat nongeneric for $n\geq 2$, it allows tree-level stabilization to be carried out in a very simple and explicit manner. Our main computational tool is the framework developed in \cite{D&DI} for abstractly describing vacua independent of a choice of Calabi-Yau threefold, in terms of $2n+2$ universal ``flux-modulus'' variables (our coinage) that are built from the superpotential and its derivatives. Adopting this formalism, we show that, for all $n=1,2,3$, both supersymmetric and nonsupersymmetric vacua with various adjustable physical parameters can be realized at almost any point in the large-complex-structure, weak-coupling moduli space of a given compactification.

In order to evaluate the true utility and relevance of these tree-level models, our field of view must necessarily extend to the stabilization of K\"ahler moduli and the effects of stringy ($\alpha'$ and $g_s$) corrections. To this end, we perform a simple but general analysis, which may be interesting in its own right, of how perturbative corrections to the ${\cal N}=1$ K\"ahler potential and nonperturbative corrections to the superpotential propagate to and affect quantities like the scalar potential and its derivatives. This analysis extends similar treatments in \cite{BB} and \cite{CQS}, but focuses on the axion-dilaton/complex-structure sector. We derive several conditions that tree-level vacua must satisfy to allow K\"ahler-modulus stabilization, a consistent cosmological constant, low-scale (hidden-sector) supersymmetry breaking, and stability beyond tree level. Along the way, we (re)classify classic stabilization scenarios such as KKLT \cite{KKLT} and the large-volume nonsupersymmetric compactifications of \cite{BBCQ}.

Non-supersymmetric tree-level solutions are often disfavored in the literature. The natural scale of supersymmetry breaking in the complex structure sector is so high that the resulting vacua often attain vacuum energies comparable to the string scale, thereby invalidating the effective field theory approach to vacuum stabilization. In principle, acceptably low supersymmetry-breaking scales could be attained at tree level by tuning fluxes, but there then appears a considerable risk that stringy corrections will become significant, causing one to lose any tree-level control. Nevertheless, through our analysis of corrections, we find that there \emph{does} exist an intermediate regime of supersymmetry breaking in the axion-dilaton/complex structure sector that could lead to controllable, consistent nonsupersymmetric solutions. We show how this scenario can be realized in our large-complex-structure models. As described in \cite{SS}, tree-level supersymmetry breaking can offer a convenient way to ``uplift'' the cosmological constant to positive values.

 Not surprisingly, we find (or reaffirm) that supersymmetric tree-level vacua can also form a basis for controllable, consistent stabilizations. We explain how they may be realized in our models as well.

Our work was initially motivated by a claim in \cite{D&DI}, that there are no stable tree-level $n=1$ supersymmetry-breaking vacua at large complex structure. This claim does not contradict the stable tree-level $n=1,2,3$ models discussed here, because we minimize the ``no-scale'' supergravity potential $V\sim|D_iW|^2$, with $i$ summing over axion-dilaton and complex-structure moduli only, whereas \cite{D&DI} (and similar statistical works) minimize a potential $V'\sim |D_iW|^2-3|W|^2$. The no-scale potential is well known to be the appropriate effective potential at tree level \cite{GKP}, and is typically more stable than $V'$ even for nonsupersymmetric vacua. Intriguingly, we find from our analysis of stringy corrections that selecting stable tree-level vacua of $V'$ may be roughly equivalent to selecting vacua of the no-scale potential that allow desirable physical properties like a low scale of supersymmetry breaking. This suggests a reinterpretation of the use of $V'$ in statistical calculations. We briefly comment on this matter at the end of the paper.

Note that our analysis neglects open-string moduli in order to further simplify calculations. Generically, both D3 and D7-branes are present in orientifold compactifications, so such a simplification, while not uncommon, is not immediately justified. D3-brane moduli enter the scalar potential by mixing with the K\"ahler structure \cite{DG, Grana:2003ek, Baumann:2006th}, and could be subsumed by our analysis of stringy corrections, but D7-brane moduli enter at tree level alongside complex-structure moduli \cite{GVW, Lust:2005bd, CDE}. Nevertheless, we will focus on the complex-structure moduli, essentially assuming that all D7-branes wrap rigid cycles, and hope that our results can be extended to the open-string sector in future work. \\

We begin in Section \ref{sec:formalism} by setting forth our general conventions, defining the flux-modulus variables, and discussing their most important properties. In Section \ref{sec:MODELS} we present the models of tree-level moduli stabilization at large complex structure, in the language of flux-modulus variables. In Section \ref{sec:analysis}, we analyze the effects of stringy corrections on tree-level vacua and apply the results to our models. Finally, in Section \ref{sec:scale-noscale}, we briefly consider the alternative scalar potential used by \cite{D&DI} and other statistical works.

\newpage
\section{Formalism}
\label{sec:formalism}

As described in the introduction, we wish to work with type IIB string theory compactifications to four dimensions on Calabi-Yau O3/O7 orientifolds with internal 3-form flux. These compactifications were developed in \cite{GKP, DG} and have been reviewed extensively, \eg\ in \cite{Denef,Grana:2005jc,Douglas:2006es}. Most of our notation is consistent with \cite{Denef}. In Section \ref{sec:conv} we recall some basic facts about the compactifications and their resulting low-energy effective theories, while explaining our general conventions. We will focus for the moment mainly on tree-level structure; further corrections will be discussed in Section \ref{sec:analysis}. In Sections \ref{sec:D&D} and \ref{sec:phys}, we introduce the flux-modulus variables of \cite{D&DI} and describe their great advantages in analyzing tree-level modulus stabilization and the physical properties of vacua.

\subsection{General conventions and notation}
\label{sec:conv}

Let $Y$ be the Calabi-Yau threefold that, after orientifolding, forms the compact space in our models. We will always assume that the expected volume of $Y$ is large in string units,
\be R^6 \equiv \langle\mbox{Vol}(Y)\rangle\gg\alpha'^3 \,, \ee
putting the Kaluza-Klein scale well below the string scale and providing a natural small parameter ($\alpha'/R^2$) for perturbation theory. The resulting $\alpha'$ expansion of interesting quantities like the four-dimensional K\"ahler potential subsumes the $g_s$ expansion, since string loop effects are only thought to contribute at higher order in $\alpha'$ \cite{Berg:2005ja, BHP}. Thus by ``tree level'', referring to $\alpha'$ tree level, we also mean tree level in $g_s$.

The orientifold action on $Y$ is generated by a holomorphic involution $\sigma$ satisfying $\sigma^2=1$, under which the holomorphic 3-form of $Y$ is odd, $\sigma^*\Omega = -\Omega$; correspondingly, the discrete symmetry $(-1)^F\sigma\, \Omega_P$ is gauged in the string theory (see \eg \cite{Sen-FO}). Under $\sigma$, the Dolbeault cohomology classes of $Y$ and respective Hodge numbers split into even and odd parts, denoted by a subscript $\pm$. The four-dimensional effective theory resulting from compactification on $Y/\sigma$ is then ${\cal N}=1$ supergravity with $h^{(1,1)}(Y)+h^{(2,1)}_-(Y)+1$ chiral multiplets, whose scalar components correspond to $h^{(1,1)}(Y)$ complexified K\"ahler moduli (and two-form axions%
\footnote{Strictly speaking, only $h^{(1,1)}_+(Y)$ K\"ahler moduli survive the orientifold projection, while another $h^{(1,1)}_-(Y)$ chiral multiplets of the the ${\cal N}=1$ theory come from axions of the 2-forms $B_2$ and $C_2$.  We will not say too much about the axions here, since for constant axion-dilaton they do not spoil the no-scale structure of the K\"ahler potential \cite{GL}, their stabilization is not required for cosmological consistency, and one can also find Calabi-Yau orientifolds with $h^{(1,1)}_-(Y)=0$ where they do not appear at all (see \emph{e.g.} the discussion in \cite{Denef}).}%
) $\rho^\alpha$,\, $n\equiv h^{(2,1)}_-(Y)$ complex moduli $t^a$, and one axion-dilaton modulus $\tau$ \cite{Grana:2003ek, GL}. Note that $h^{(3,0)}_-(Y)=1$ and $h^{(3,0)}_+(Y)=0$. Only the $\sigma$-odd part of the middle cohomology is relevant after the orientifold projection. We will often use an index $i=0,...,n$ to describe both complex-structure moduli and the axion-dilaton (with $i=0$ denoting the axion-dilaton), and capital indices $A,B,...$ to describe all moduli together:
\be \begin{array}{c|ccc}
 \mbox{moduli} \,& \multicolumn{3}{c}{\mbox{indices}}  \\ \hline
 \mbox{axion-dilaton} \,&\, 0 & \multirow{2}{*}{$\bigg\}\;i,j=0,...,n$} & \multirow{3}{*}{\Bigg\} $A,B$} \\
 \mbox{complex structure} \,&\, a,b = 1,...,n & &\\
 \mbox{K\"ahler structure} \,&\, \alpha,\beta & &
 \end{array} \label{indices}
\ee

We work in conventions where all the moduli fields are dimensionless. The imaginary parts of the $\rho^\alpha$ are four-cycle volumes measured in string units, and the axion-dilaton is
\be \tau = C_0 + ie^{-\phi} = C_0 + i/g_s \,. \ee
Given a symplectic basis $\{A_i,B^i\}_{i=0}^n$ of $H_3^{-}(Y)$, Poincar\'e dual to a symplectic basis for odd middle cohomology, homogeneous special coordinates on the moduli space of complex structures ${\cal M}_\mbc(Y)$ surviving the projection are given by
\be w^i = \int_{A_i}\Omega \,, \quad i=0,...,n \,, \label{periods0} \ee
and the complex-structure moduli are defined as
\be t^a = w^a/w^0 \,, \quad a=1,...,n \,. \ee
In other words, the $t^a$ correspond to a gauge (normalization of $\Omega$) such that $w^0=1$. If $\{A_i,B^i\}$ is an integral basis of $H_{3}^{-}(Y;\mbz)$, then the $w^i$ are unique up to $Sp(2n+2;\mbz)$ transformations. More generally, $\{A_i,B^i\}$ can simply be a real basis of $H_3^-(Y,\mbr)$, and then the $w^i$ are unique up to $Sp(2n+2;\mbr)$ \cite{Str-SG}.

We normalize the 3-form fluxes by $1/(2\pi)^2\alpha'\sim\sqrt{T_3}$, where $T_3$ is the Einstein-frame D3-brane tension, so that their integrals are dimensionless and they are represented directly by integral cohomology. The complexified, $SL(2,\mbz)$-invariant flux $G_3$ is then defined as
\be G_3 = F_{3} - \tau H_{3}\,, \qquad F_{3} = dC_2\,,\quad H_{3}=dB_2\,, \ee
\ben  F_{3},\,H_{3}\in H^{3}_-(Y;\mbz)\,, \een
where $F_3$ is the internal RR flux and $H_3$ is the internal NSNS flux.  At tree level, we can neglect any conformal warping of $Y$ due to backreaction of the fluxes \cite{DG, CQS}.

The four-dimensional effective potential for the scalar moduli is given in ${\cal N}=1$ supergravity by
\be  V = T_3\, e^K
 (g^{A\bar{B}}D_AW\bar{D}_{\bar{B}}\bar{W}-3|W|^2)\,, \label{V} \ee
up to an $O(1)$ constant prefactor. At tree-level, the flux-generated superpotential $W$ is \cite{GVW, GKP}
\be W = \int_Y G_3\wedge\Omega\,, \label{W0} \ee
and the K\"ahler potential is
\begin{align} K &= -\log[-i(\tau-\bar{\tau})]-\log i \int_Y\Omega\wedge\bar{\Omega} - 2\log {\cal V} \label{K0} \\
 &\equiv K_\tau + K_\mbc + K_K\,, \end{align}
splitting into axion-dilaton, complex-structure, and K\"ahler pieces. The quantity ${\cal V}$, essentially%
\footnote{The form of \eqref{V}, with a $T_3$ prefactor, already includes a $[(\alpha')^3/\langle\mbox{Vol}(Y)\rangle]^2$ contribution that would have come from the Calabi-Yau volume in $e^{K_K}$. To be consistent, the $\cal V$ in \eqref{K0} should really only measure \emph{fluctuations} around the expected large volume $R^6$, in string units. For a careful dimensional reduction of the 10-dimensional supergravity action including proper units, see \cite{DG}. \label{foot:vol}} %
the volume of $Y$, contains the dependence on K\"ahler moduli. In our conventions, both $K$ and $W$ are dimensionless; the physical supergravity expressions are given by
\be K^{sugra} = M_P^2K\,,\qquad W^{sugra} = M_P\sqrt{T_3}W\,, \label{sugra_pot} \ee
where
\be M_P^2 \sim \frac{\langle\mbox{Vol}(Y)\rangle}{\alpha'^4} \ee
is the 4-dimensional Planck mass. The metric $g^{A\bar{B}}$ is the inverse of the K\"ahler metric $g_{\bar{A}B}=\pd_{\bar{A}}\pd_BK$ on the total moduli space ${\cal M} = {\cal M}_\tau\times {\cal M}_\mbc(Y)\times {\cal M}_K(Y)$.

The K\"ahler (and metric) covariant derivative acts as $D_AW = \pd_AW+(\pd_AK) W$ on $W$, since $W$ is a scalar section of the holomorphic K\"ahler line bundle ${\cal L}$ on $\cal M$. For an arbitrary section $\varphi$ of ${\cal L}^h\otimes\bar{\cal L}^{\bar{h}}$, one defines 
\be D_A \varphi = \nabla_A \varphi + h(\pd_AK)\varphi + \bar{h}\ol{(\pd_AK)}\varphi\,, \ee
with $\nabla_A$ being the usual metric connection \cite{Str-SG, Candelas:1990pi}. This covariant derivative always commutes with powers of $e^K \in \Gamma({\cal L}^{-1}\otimes\bar{\cal L}^{-1})$, provided one keeps track of K\"ahler weights.

Because the superpotential \eqref{W0} is independent of K\"ahler moduli, and the K\"ahler potential \eqref{K0} satisfies $g^{\alpha\bar{\alpha}}\pd_\alpha K \pd_{\bar{\alpha}}\bar{K} = 3$, the tree-level scalar potential assumes the standard no-scale form
\be V = (T_3)\,e^Kg^{i\bar{j}}D_iW\bar{D}_{\bar{j}}\bar{W}\,, \label{V-noscale} \ee
summing over axion-dilaton/complex directions only. There is no dependence on K\"ahler moduli aside from a prefactor $\sim 1/{\cal V}^2$ coming from $e^K$. This is the potential we will use in our models; as observed in (\eg) \cite{SS,Becker:2007ee}, it can have (meta)stable supersymmetry-breaking minima in the axion-dilaton and complex-structure directions in addition to the supersymmetric minimum at $D_iW=V=0$.

\subsection{Flux-modulus variables}
\label{sec:D&D}

Following \cite{D&DI} (see also \cite{Kallosh:2005ax, Bellucci:2007ds}), we introduce mixed flux-modulus variables that greatly simplify the analysis of tree-level vacua. Rescaling the superpotential as $\tilde{W} = e^{K/2}W$, these $2n+2$ complex variables $(X,Y_i,Z_a)$ are defined by
\be \fbox{$
\begin{array}{rcl}
  X &=& \tilde{W} \\
 Y_i &=& D_i\tilde{W} \\
 Z_a &=& D_0 D_a\tilde{W}.
\end{array}
$} \label{XYZ} \ee
We call them ``flux-modulus'' variables because they also arise as coefficients of the flux form $G_3$ expanded in a Hodge basis of $H^3(Y)\otimes\mbc$, an expansion which depends on the axion-dilaton and complex-structure moduli.

It is convenient to work in an orthonormal (really, unitary-normal) frame on $T{\cal M} = T{\cal M}_\tau \oplus T{\cal M}_\mbc \oplus T{\cal M}_K$, defined by the vielbeins $e{^A}_{\hat{B}} = e{^0}_{\hat{0}}\oplus e{^a}_{\hat{b}}\oplus e{^\alpha}_{\hat{\beta}}$. We denote orthonormal-frame tensors with hatted indices. Then, as discussed in \cite{D&DI}, further covariant derivatives of the tree-level superpotential only depend algebraically on $(X,Y_i,Z_a)$ as well as the rescaled Yukawa couplings
\be \cF_{abc} \equiv ie^{K_\mbc} \int_Y \Omega \wedge D_aD_bD_c\Omega = ie^{K_\mbc} \int_Y \Omega \wedge \pd_a\pd_b\pd_c \Omega\,. \ee
In particular,
\begin{subequations} \label{DWs}
\begin{align} D_{\hat{0}}D_{\hat{0}} \tilde{W} &= 0 \\
D_{\hat{a}} D_{\hat{b}} \tilde{W} &= \cF_{\hat{a}\hat{b}\hat{c}}\bar{Z}^{\hat{c}} \\
D_{\hat{0}}D_{\hat{a}}D_{\hat{b}} \tilde{W} &=  \cF_{\hat{a}\hat{b}\hat{c}}\bar{Y}^{\hat{c}} \\
D_{\hat{a}}D_{\hat{b}}D_{\hat{c}} \tilde{W} &=  \,D_{\hat{a}}\cF_{\hat{b}\hat{c}\hat{d}}\bar{Z}^{\hat{d}} +  \cF_{\hat{a}\hat{b}\hat{c}}\bar{Y}^{\hat{0}}.
\end{align}
\end{subequations}

The first and second covariant derivatives of the full potential \eqref{V} (setting $T_3\goesto 1$) can be written generically as
\begin{subequations} \label{DVgen}
\begin{align} e^{-K}V &= g^{C\bar{D}}D_CW\bar{D}_{\bar{D}}\bar{W}-3 W\bar{W} \\
e^{-K}D_AV &= g^{C\bar{D}}D_AD_CW \bar{D}_{\bar{D}}\bar{W} + (1-3)D_AW \bar{W} \\
e^{-K}D_AD_BV &= g^{C\bar{D}}D_AD_BD_CW\bar{D}_{\bar{D}}\bar{W}+(2-3)D_AD_BW \bar{W} \\
e^{-K}\bar{D}_{\bar{A}}D_BV &= g^{C\bar{D}}D_BD_CW \bar{D}_{\bar{A}}\bar{D}_{\bar{D}}\bar{W} - {R_{\bar{A}B}}^{C\bar{D}}D_CW\bar{D}_{\bar{D}}\bar{W} + g_{\bar{A}B} g^{C\bar{D}}D_CW\bar{D}_{\bar{D}}\bar{W} \nno \\
&\quad + (2-3)D_BW\bar{D}_{\bar{A}}\bar{W}+(1-3)g_{\bar{A}B}W\bar{W},
\end{align}
\end{subequations}
with the Riemann tensor defined via ${R_{A\bar{B}C}}^DT_D \equiv [\nabla_A,\bar{\nabla}_{\bar{B}}]T_C$. At tree level, the no-scale relation effectively restricts internal summations to $i,j$ indices while removing contributions from the $-3|W|^2$ term and its derivatives, which have been explicitly tracked above. Then, using the fact that ${R_{0\bar{0}}}^{\bar{0}0}=2$ as well as the special geometry identity $R_{a\bar{b}c\bar{d}} = -g^{e\bar{e}}\cF_{ace}{\bar{\cF}}_{\bar{b}\bar{d}\bar{e}}+g_{a\bar{b}}g_{c\bar{d}}+g_{a\bar{d}}g_{c\bar{b}}$ (or via direct differentiation), the tree-level derivatives in complex-structure and axion-dilaton directions can be written in terms of flux-modulus variables as
\begin{subequations} \label{derV0}
\begin{align}
 V &=  Y_{\hat{i}}\bar{Y}^{\hat{i}} \\
 D_{\hat{0}} V &= Z_{\hat{c}} \bar{Y}^{\hat{c}} + Y_{\hat{0}}\bar{X} \\
 D_{\hat{a}} V &= Z_{\hat{a}}\bar{Y}_{\hat{0}} + \cF_{\hat{a}\hat{b}\hat{c}}\bar{Z}^{\hat{b}}\bar{Y}^{\hat{c}}+Y_{\hat{a}}\bar{X} \\
 D_{\hat{0}}D_{\hat{0}}V &= 0 \label{firstmass} \\
 D_{\hat{0}}D_{\hat{a}}V &= \cF_{\hat{a}\hat{b}\hat{c}}\bar{Y}^{\hat{b}}\bar{Y}^{\hat{c}}+2Z_{\hat{a}}\bar{X} \\
 D_{\hat{a}}D_{\hat{b}}V &= 2\cF_{\hat{a}\hat{b}\hat{c}}\bar{Y}^{\hat{0}}\bar{Y}^{\hat{c}} + 2\cF_{\hat{a}\hat{b}\hat{c}}\bar{Z}^{\hat{c}} \bar{X} + \framebox{$D_{\hat{a}}\cF_{\hat{b}\hat{c}\hat{d}}$}\bar{Y}^{\hat{c}}\bar{Z}^{\hat{d}} \label{DabV} \\
 \bar{D}_{\hat{\bar{0}}}D_{\hat{0}}V &= |X|^2 + Y_{\hat{i}}\bar{Y}^{\hat{i}} + Z_{\hat{c}}\bar{Z}^{\hat{c}} \\
 \bar{D}_{\hat{\bar{0}}}D_{\hat{a}}V &= 2 \bar{Y}_{\bar{{\hat{0}}}}Y_{\hat{a}} + \cF_{\hat{a}\hat{b}\hat{c}}\bar{Z}^{\hat{b}}\bar{Z}^{\hat{c}} \\
 \bar{D}_{\hat{\bar{a}}}D_{\hat{b}}V &= \delta_{\hat{\bar{a}}\hat{b}}(|X|^2+Y_{\hat{0}}\bar{Y}^{\hat{0}})+\bar{Y}_{\bar{a}}Y_{\hat{b}}+\bar{Z}_{\hat{\bar{a}}}Z_{\hat{b}}+\bar{\cF}_{\hat{\bar{a}}\hat{\bar{c}}\hat{\bar{e}}}{\cF^{\hat{\bar{e}}}}_{{\hat{b}}{\hat{d}}}(Y^{\hat{\bar{c}}}\bar{Y}^{\hat{d}}+Z^{\hat{\bar{c}}}\bar{Z}^{\hat{d}}) \,. \label{lastmass}
\end{align}
\end{subequations}
The only appearance of a derivative of the Yukawa couplings is in \eqref{DabV}.

Note that we have defined flux-modulus variables in terms of a superpotential rescaled by the \emph{full} K\"ahler potential $e^{(K_\tau+K_\mbc+K_K)/2}$. The $e^{K_K/2}$ rescaling is simply a formality, which makes many expressions look cleaner but gives $(X,Y_i,Z_a)$ a pseudo-dependence on K\"ahler moduli, which are unfixed at tree level. Since all quantities of interest (in particular $V$ and its covariant derivatives) depend homogeneously on flux-modulus variables, we could always move all $e^{K_K/2}$ rescalings into prefactors.

\subsection{Physical properties}
\label{sec:phys}

The potential advantage of flux-modulus variables is that they ``factorize'' the problem of stabilization in the axion-dilaton/complex-structure sector. Acting on the scalar $V$, covariant derivatives are equivalent to ordinary derivatives. From \eqref{derV0}, one sees that the critical point equations $\pd_i V = D_iV = 0$ only involve $(X,Y_i,Z_a)$ and the Yukawa couplings (and their conjugates) algebraically; thus, for any given $n$, all possible solutions to $\pd_iV=0$ can be described by some abstract, universal algebraic variety ${\cal X}_n$. The geometric data of a particular Calabi-Yau orientifold is encoded in the dependence of $(X,Y_i,Z_a;\cF_{abc})$ on moduli and fluxes, and stabilization is achieved when moduli and/or fluxes are tuned so that $(X,Y_i,Z_a;\cF_{abc}) \in {\cal X}_n$.

Many physically interesting tree-level quantities depend only on the flux-modulus variables and Yukawa couplings rather than a specific choice of Calabi-Yau, and in this sense can be analyzed universally. This was of course the motivation behind using these variables in the statistical analyses of \cite{D&DI, D&DII}. Most basically, the cosmological constant associated to a certain tree-level vacuum is
\be \Lambda = V|_{vac} = T_3|\Y{i}|^2 \sim \frac{1}{(\alpha')^2}|\Y{i}|^2\,, \ee
and the tree-level scale of supersymmetry breaking and gravitino mass are
\be \frac{M_{susy}^4}{M_P^4} = \frac{T_3}{M_P^4}|\D{i}\tW|^2 \sim \left(\frac{\alpha'}{R^2}\right)^6|Y_{\hat{i}}|^2\,, \qquad \frac{M_{3/2}^2}{M_P^2} = \frac{T_3}{M_P^4}|\tW|^2 \sim \left(\frac{\alpha'}{R^2}\right)^6|X|^2\,. \label{susygrav} \ee
Also, the chiral mass matrix for the modulinos, the fermionic superpartners of complex-structure and axion-dilaton moduli, is given in supergravity by $(M_{\tilde{\mbc},\tilde{\tau}})_{ij} \sim \frac{\sqrt{T_3}}{M_P}D_iD_j\tilde{W}$ \cite{WB}. The extra factors $\sqrt{T_3}$ and $1/M_P$ enter from the physical K\"ahler and superpotentials as in \eqref{sugra_pot}. In an orthonormal frame, which is most appropriate for considering physical masses, the components of $M_{\tilde{\mbc},\tilde{\tau}}$ are just
\begin{subequations} \label{modulinos}
\begin{align} (M_{\tilde{\mbc},\tilde{\tau}})_{\hat{0}\hat{0}}&=0\,, \\
 (M_{\tilde{\mbc},\tilde{\tau}})_{\hat{0}\hat{a}}&\sim\frac{\alpha'}{R^3}\Z{a}\,, \\
 (M_{\tilde{\mbc},\tilde{\tau}})_{\hat{a}\hat{b}}&\sim\frac{\alpha'}{R_3}\cF_{\hat{a}\hat{b}\hat{c}}\bar{Z}^{\hat{c}}\,.
\end{align}
\end{subequations}
The RHS are of course to be evaluated at the values of the moduli in a given stable vacuum.

The masses of the moduli themselves are obtained by diagonalizing the mass matrix $M_P^{-2}\pd\bar{\pd}V$. Again, the extra factor of $M_P^{-2}$ enters from the definition of the physical K\"ahler potential. Since $\pd_j\pd_k V = D_j D_k V$ when $\pd_iV=0$, the entries of the tree-level mass matrix can all be obtained from (\ref{derV0}d-i). Indeed, as a consequence of the no-scale cancellation, we find that the mass matrix has a very convenient decomposition. In an orthonormal frame,
\be M^2_{\mbc,\tau} = \left(\begin{array}{cc} 
 \bar{D}^{\hat{i}}D_{\hat{j}}V & \bar{D}^{\hat{i}}\bar{D}_{\hat{\bar{j}}}V \\
 D^{\hat{\bar{i}}}D_{\hat{j}}V & D^{\hat{\bar{i}}}D_{\hat{\bar{j}}}V
 \end{array}\right) = H^2 + A^2 + d\,, \label{M-decomp}
\ee
with
\begin{subequations} \label{M-decomp2}
\be H = \frac{1}{|X|}\left(\begin{array}{cccc}
 |X|^2 & 0 & 0 & X\bar{Z}_{\hat{\bar{b}}} \\
 0 & |X|^2 {\delta^{\hat{a}}}_{\hat{b}} & X \bar{Z}^{\hat{a}} & X(\bar{\cF}\ccdot Z){^{\hat{a}}}_{\hat{\bar{b}}} \\
 0 & \bar{X}Z_{\hat{b}} & |X|^2 & 0 \\
 \bar{X}Z^{\hat{\bar{a}}} & \bar{X}(\cF\ccdot\bar{Z}){^{\hat{\bar{a}}}}_{\hat{b}} & 0 & |X|^2\delta{^{\hat{\bar{a}}}}_{\hat{\bar{b}}}
 \end{array}\right), \ee
\be A = \frac{1}{|Y_{\hat{0}}|}\left(\begin{array}{cccc}
 |Y_{\hat{0}}|^2 & \bar{Y}^{\hat{0}}Y_{\hat{b}} & 0 & 0 \\
 Y^{\hat{\bar{0}}}\bar{Y}^{\hat{a}} & |Y_{\hat{0}}|^2\delta{^{\hat{a}}}_{\hat{b}} & 0 & Y^{\hat{\bar{0}}}(\bar{\cF}\ccdot Y){^{\hat{a}}}_{\hat{\bar{b}}} \\
 0 & 0 & |Y_{\hat{0}}|^2 & Y^{\hat{\bar{0}}}\bar{Y}_{\hat{\bar{b}}} \\
 0 & \bar{Y}^{\hat{0}}(\cF\ccdot\bar{Y}){^{\hat{\bar{a}}}}_{\hat{b}} & \bar{Y}^{\hat{0}}Y^{\hat{\bar{a}}} & |Y_{\hat{0}}|^2\delta{^{\hat{\bar{a}}}}_{\hat{\bar{b}}}
 \end{array}\right), \ee
\be d = \left(\begin{array}{cccc}
 0 & 0 & 0 & 0 \\
 0 & 0 & 0 & (\bDFYZ){^{\hat{a}}}_{\hat{\bar{b}}} \\
 0 & 0 & 0 & 0 \\
 0 & (\DFYZ){^{\hat{\bar{a}}}}_{\hat{b}} & 0 & 0
 \end{array}\right), \ee
where
\be (\cF\ccdot\bar{Z})_{\hat{a}\hat{b}} = \cF_{\hat{a}\hat{b}\hat{c}}\bar{Z}^{\hat{c}}\,, \qquad
    (\cF\ccdot\bar{Y})_{\hat{a}\hat{b}} = \cF_{\hat{a}\hat{b}\hat{c}}\bar{Y}^{\hat{c}}\,, \qquad
    (\DFYZ)_{\hat{a}\hat{b}} = D_{\hat{a}}\cF_{\hat{b}\hat{c}\hat{d}}\bar{Y}^{\hat{c}}\bar{Z}^d\,. \nno \ee
\end{subequations}
This shows that stability (positive-definiteness) at tree level is \emph{almost} guaranteed,%
\footnote{A similar decomposition of the mass matrix for the no-scale potential appears in \cite{Becker:2007ee}, but neglects the instabilities arising from $D\cF\neq 0$.} %
 and offset only by the off-diagonal matrix $d$. In particular, supersymmetric vacua $(\Y{i}\equiv0)$, vacua with $\Z{a}\goesto0$, and vacua with $\DFab\goesto0$ are all automatically stable.
\footnote{Throughout this paper we will only consider stability of vacua in a local sense. Many of the supersymmetry-breaking vacua that we analyze could arise from potentials containing other, lower-energy vacua (\eg\ supersymmetric vacua), but we neglect here the possibility of tunneling.}

The D3 tadpole condition for O3/O7 orientifolds also assumes a particularly nice form in flux-modulus variables. Generally one must have
\be N_{flux} + N_{D3}-\frac{1}{4}N_{O3}-\frac{\chi(X)}{24} = 0\,, \label{pre-tadpole} \ee
where $N_{D3}$ and $N_{O3}$ count the numbers of D3 branes and O3 planes, $\chi(X)/24$ is a contribution from $D7$ branes and $O7$ planes (written in terms of the F-theory fourfold $X$ whose orientifold limit produces $Y$), and $N_{flux}$ is the flux contribution \cite{GKP, GVW}
\be N_{flux} = \int_Y F_3\wedge H_3 \in \mbz\,. \ee
In flux-modulus variables, the flux contribution is just
\be N_{flux} = |X|^2-|Y_{\hat{i}}|^2+|Z_{\hat{a}}|^2\,. \label{tadpole} \ee
It is well-known that the flux contribution is only positive-definite for supersymmetric vacua (\ie\ those with $Y_{\hat{i}}\equiv 0$) \cite{D0, AD, Denef}; this is easily seen in \eqref{tadpole}. Typically, the $-\chi(X)/24$ contribution to \eqref{pre-tadpole} is large and negative, so $N_{D3}$ can be adjusted to satisfy \eqref{pre-tadpole} for any $N_{flux}$ within a (large) given range; thus we will not worry about cancelling the tadpole in our toy models, at their current level of detail.

\section{Models at large complex structure}
\label{sec:MODELS}

We now present our models of tree-level moduli stabilization near large complex structure, using the flux-modulus formalism of the preceding sections. We compactify on a special class of Calabi-Yau threefolds $Y$ with $n=1,\,2,$ and $3$ complex-structure moduli (after orientifold projection), which have a single nonvanishing Yukawa coupling in the large-complex-structure limit. This special property ensures that the orthonormal-frame Yukawa couplings are both constant and covariantly constant near large complex structure, and allows the ``factorization'' of moduli stabilization described in Section \ref{sec:phys} to be realized very explicitly. In particular, (1) we are able to describe the set of all possible solutions to $dV=0$ (\ie\ the varieties ${\cal X}_n$) abstractly in terms of flux-modulus variables alone, with no dependence on Yukawa couplings; and (2) for any particular Calabi-Yau, we can actually produce a vacuum corresponding to any abstract $(X,\Y{i},\Z{a})\in{\cal X}_n$ at any chosen values of moduli $(\tau, t^a)$ by simply tuning the internal fluxes. As seen from the decomposition of the mass matrix $M^2_{\tau,\mbc}$ in \eqref{M-decomp}, an added benefit of covariantly constant Yukawa couplings is that moduli masses are always positive. Thus, every abstract solution to $dV=0$ really does correspond to a (meta)\emph{stable} tree-level vacuum.

We will begin in Section \ref{sec:family} by making the single-Yukawa condition more precise and deriving the main properties of the resulting models. We will also attempt to quantify just how restrictive the single-Yukawa condition actually is. In Section \ref{sec:maps}, we will justify the second claim above, that any desired abstract vacuum can actually be attained by tuning fluxes. We explicitly show the dependence of flux-modulus variables on fluxes and moduli in the simplest case $n=1$, deferring further details of $n=2$ and $n=3$ to Appendix \ref{app:mapdata}. We treat fluxes as continuous variables throughout most of this analysis, but comment on the effects of flux quantization at the end of Section \ref{sec:maps}. In Section \ref{sec:X}, we finally describe the abstract varieties ${\cal X}_{1}$ and ${\cal X}_{2}$ for our models in terms of flux-modulus variables, thereby classifying all the possible vacua. We defer the case $n=3$ (${\cal X}_3$) to Appendix \ref{app:n3sol}.

As noted in the introduction, the true utility of these models depends on their ability to form consistent stabilizations beyond tree level, incorporating $\alpha'$-corrections and the stabilization of K\"ahler moduli. Therefore, we delay a full discussion of the models' physical properties (as predicted by the varieties ${\cal X}_n$) until Section \ref{sec:modelanal}.

\subsection{A well-behaved family of Calabi-Yau orientifolds}
\label{sec:family}

Recall that in general the prepotential on ${\cal M}_\mbc(Y)$ is constructed from the periods of $\Omega$ as $\cF = \frac{1}{2}w^i F_i$, where $w^i = \int_{A_i}\Omega$ as in \eqref{periods0} and $F^i = \int_{B^i}\Omega$, and that it takes the form \cite{Candelas:1990rm, Hosono:1994av}
\begin{align} F &= -\frac{1}{6}y_{abc}\frac{w^aw^bw^c}{w^0} - \frac{1}{2}q_{ab}w^aw^b - \ell_a w^a w^0 - c (w^0)^2 + F_{inst} \nno \\
 &= -\frac{1}{6}y_{abc}t^at^bt^c - \frac{1}{2}q_{ab}t^at^b-\ell_at^a-c + F_{inst}\,. \label{prep}
\end{align}
The term $F_{inst}$ contains contributions from worldsheet instantons $\sim e^{2\pi i t^a}$, which can be neglected in the large complex structure limit, defined as
\be \Im\,t_a \goesto \infty\,. \ee
The real, symmetric, constant tensor $y_{abc}$ is then related to our rescaled Yukawa couplings via $\cF_{abc} = -ie^{K}y_{abc}$. The precise restriction we impose on our compactification manifolds is that all but one component of $y_{abc}$, up to symmetry, vanishes, for some choice of special coordinates.

Before investigating the properties of the resulting models, let us ask just how generic they are. The single-Yukawa restriction is only reasonable when $n\leq 3$, because the indices of the nonvanishing coupling must involve all the complex-structure moduli. Otherwise, some modulus will not appear in the the K\"ahler potential (\emph{cf}. \eqref{large_K} below), leading to a degenerate metric in our large-complex-structure approximation. We can then consider each case $n=1,2,3$ separately.

For $n=1$, there is a unique Yukawa coupling $y_{111}$, so the restriction is satisfied automatically. For $n=2$, however, there are four distinct Yukawa couplings up to permutations of the indices. Suppose we want only $y_{112}$ to be nonvanishing (we cannot choose $y_{111}$ or $y_{222}$ to be nonvanishing because every index must appear, and choosing $y_{122}$ is equivalent to choosing $y_{112}$). A subset of special coordinate transformations can effect $SL(2,\mbr)$ transformations on $(t^1,t^2)$. A generic polynomial $\frac{1}{6}y_{abc}t^at^bt^c$ can then be transformed to the desired form $\frac{1}{2}y_{112}'{t^1}'{t^1}'{t^2}'$ if it has a double root but not a triple root. In other words, there exist special coordinates in which only $y_{112}'$ is nonvanishing as long as the discriminant
\be \Delta = 3 y_{112}^2y_{122}^2 - 4y_{112}^3 y_{222}-4y_{111}y_{122}^3-y_{111}^2y_{222}^2 + 6y_{111}y_{112}y_{122}y_{222} \label{disc} \ee
vanishes, and either
\be y_{112}^2\neq y_{111}y_{122} \qquad\mbox{or}\qquad y_{112}^3\neq y_{111}^2 y_{222}\,, \ee
to prevent the triple root. Thus, we can think of our restricted models as comprising a subset of codimension $1$ in the space of possible Yukawa couplings. For $n=3$ the situation is similar. There are ten distinct couplings, and requiring the polynomial $\frac{1}{6}y_{abc}t^at^bt^c$ to become $y_{123}'{t^1}'{t^2}'{t^3}'$ under a change of coordinates restricts the couplings to a subset of codimension 3.

Note that two of the best-studied Calabi-Yau manifolds, the mirrors of the degree-eight hypersurfaces in $\mathbb{P}_4^{(1,1,2,2,2)}$ and the degree-twelve hypersurfaces in $\mathbb{P}_4^{(1,1,2,2,6)}$, belong to our family of models with $n=2$. In the original notation of \cite{C-MSI}, these manifolds have $y_{122}=y_{222}=0$ due to a nilpotent divisor, so they trivially satisfy $\Delta = 0$ (see also \cite{Misra:2004ky, Kaura:2006mv, Conlon:2004ds, CCQ}). Toroidal orbifolds preserving a product structure $T^6 = T^2\times T^2\times T^2$, discussed in \cite{D&DII} and many other places, have a single nonvanishing coupling $y_{123}\neq 0$, and so provide a somewhat special example of our family of models with $n=3$.

Now consider the geometric properties of single-Yukawa manifolds at large complex structure. In the prepotential \eqref{prep}, the constants $y_{abc}$, $q_{ab}$ and $\ell_a$ must all be real, whereas $c$ is purely imaginary and proportional to the Euler number of $Y$ \cite{Hosono:1994av}. Near large complex structure, we can neglect the instanton contribution $F_{inst}$ and compute the periods 
\begin{subequations} \label{periodsF}
\begin{align} F_0 &= \int_{B^0}\Omega = \frac{\pd}{\pd w^0}F = \frac{1}{6}y_{abc}t^at^bt^c-\ell_at^a-2c\,, \\
F_a &= \int_{B^a}\Omega = \frac{\pd}{\pd w^a}F = -\frac{1}{2}y_{abc}t^bt^c-q_{ab}t^b-\ell_a\,,
\end{align}
\end{subequations}
which lead to a complex-structure K\"ahler potential
\begin{align} K_\mbc &= -\log\left[i\int_Y\Omega\wedge\bar{\Omega}\right]  \nno \\ &
 = -\log\left[i(\bar{w}^iF_i-w^i\bar{F}_i)\right]
   \nno\\
 &= -\log\left[i\frac{1}{6}y_{abc}(t^a-\bar{t}^a)(t^b-\bar{t}^b)(t^c-\bar{t}^c)-4ic\right]. \label{large_K}
\end{align}
At reasonably large complex structure, we can also neglect $c$. Specifically, defining 
\be \Im\,t \equiv \min\big\{\Im\,t^1,\Im\,t^2,...,\Im\,t^n\big\} \qquad\mbox{and}\qquad \epsilon \equiv \frac{c}{(\Im\, t)^3} \label{def_te} \ee
as measures of our proximity to the large-complex-structure point, we have
\begin{align} K_\mbc &= -\log\left[i\frac{1}{6}y_{abc}(t^a-\bar{t}^a)(t^b-\bar{t}^b)(t^c-\bar{t}^c)\right] + \log(1+O(\epsilon))\,. \label{large_K2} \end{align}
When there is a single nonvanishing Yukawa coupling, the resulting metric on ${\cal M}_\mbc$ will be \emph{diagonal}, with components typically of order $1/(\Im\,t)^2$, up to corrections of order $\epsilon/(\Im\,t)^2$.

We can explicitly show that the orthonormal-frame Yukawa couplings are approximately constant and covariantly constant, given a single-Yukawa condition, for each of the three cases $n=1,2,3$. For $n=1$, the K\"ahler potential is $K_\mbc = -\log[\frac{1}{6}iy_{111}(t-\bar{t})^3]$, whence
\begin{subequations} \label{n1geom}
\be \pd_1K = -\frac{3}{t-\bar{t}}\,,\qquad g_{1\bar{1}}=-\frac{3}{(t-\bar{t})^2}\,,\qquad \Gamma_{11}^1 = -\frac{2}{t-\bar{t}}\,, \ee
\be e{_{\hat{1}}}^{1} = -\frac{t-\bar{t}}{\sqrt{3}}\,. \ee
\end{subequations}
Here we denote by $\Gamma^a_{bc} = g^{a\bar{d}}\pd_bg_{\bar{d}c}$ the Christoffel connection on $T{\cal M}_\mbc$. All these expressions receive corrections of fractional order $\epsilon$ from \eqref{large_K2}. The rescaled orthonormal-frame Yukawa coupling is then\footnote{This result also appears in \cite{D&DI}, Section 4.2.}
\be
 \framebox{$\displaystyle\cF_{\hat{1}\hat{1}\hat{1}} = -(e{_{\hat{1}}}^1)^3 ie^{K_\mbc}y_{111} = \frac{2}{\sqrt{3}} + O(\epsilon)$} \,. \label{n1F} \ee
Its covariant derivative is
\be D_{\hat{1}}\cF_{\hat{1}\hat{1}\hat{1}} = (e{_{\hat{1}}}^1)^4 i e^{K_\mbc}(2\pd_1 K - 3\Gamma^1_{11})y_{111} = 0 + O(\epsilon)\,, \label{D1111} \ee
vanishing to order $\epsilon$ due to a cancellation between the K\"ahler and metric connections.

For $n=2$ and (WLOG) $y_{112}\neq 0$, we similarly have $K_\mbc = -\log[\frac{1}{2}iy_{112}(t^1-\bar{t}^1)^2(t^2-\bar{t}^2)]$ and
\begin{subequations} \label{n2geom}
\be \pd_1K = -\frac{2}{t^1-\bar{t}^1} \,,\qquad \pd_2K = -\frac{1}{t^2-\bar{t}^2} \,,\qquad \Gamma_{11}^1 = -\frac{2}{t^1-\bar{t}^1}\,,\qquad
 \Gamma_{22}^2 = -\frac{2}{t^2-\bar{t}^2}\,, \ee
\be g_{a\bar{b}} = \left(\begin{array}{cc}
 -\frac{2}{(t^1-\bar{t}^1)^2} & 0 \\ 0 & -\frac{1}{(t^2-\bar{t}^2)^2}
  \end{array}\right), \qquad
  e{_{\hat{a}}}^b = \left(\begin{array}{cc}
  -\frac{t^1-\bar{t}^1}{\sqrt{2}} & 0 \\ 0 & -(t^2-\bar{t}^2)
  \end{array}\right)\,. \label{n2-K3}
\ee
\end{subequations}
All mixed components of $\Gamma^{a}_{bc}$ vanish. The only nonvanishing orthonormal-frame Yukawa coupling is
\be \framebox{$\displaystyle \cF_{\hat{1}\hat{1}\hat{2}} = -(e{_{\hat{1}}}^1)^2(e{_{\hat{2}}}^2)ie^{K_\mbc}y_{112} = 1 + O(\epsilon)$}\,, \label{n2F} \ee
and both its covariant derivatives vanish, again due to cancellations between the K\"ahler and metric connections:
\begin{align} D_{\hat{1}}\cF_{\hat{1}\hat{1}\hat{2}} &= (e{_{\hat{1}}}^1)^3(e{_{\hat{2}}}^2)ie^{K_\mbc}(2\pd_1K-2\Gamma_{11}^1)y_{112} = 0+ O(\epsilon)\,, \nno \\
 D_{\hat{2}}\cF_{\hat{1}\hat{1}\hat{2}} &= (e{_{\hat{1}}}^1)^2(e{_{\hat{2}}}^2)^2ie^{K_\mbc}(2\pd_2K-\Gamma_{22}^2)y_{112} = 0+ O(\epsilon)\,. \nno
\end{align}

For $n=3$ and $y_{123}\neq0$, the situation is more symmetric, with $K_\mbc = -\log[iy_{123}(t^1-\bar{t}^1)$ $\times(t^2-\bar{t}^2)(t^3-\bar{t}^3)]$ and
\begin{subequations} \label{n3geom}
\be \pd_aK = -\frac{1}{t^a-\bar{t}^a} \,,\qquad \Gamma^a_{aa}=-\frac{2}{t^a-\bar{t}^a} \,,\qquad g_{a\bar{a}}=-\frac{1}{(t^a-\bar{t}^a)^2}\,, \ee
\be e{_{\hat{a}}}^a = -(t^a-\bar{t}^a) \ee
\end{subequations}
for any fixed $a=1,2,3$, with all other components vanishing. The nonvanishing Yukawa coupling is
\be \framebox{$\displaystyle \cF_{\hat{1}\hat{2}\hat{3}}= -(e{_{\hat{1}}}^1)(e{_{\hat{2}}}^2)(e{_{\hat{3}}}^3)ie^{K_\mbc}y_{123} = 1+O(\epsilon)$}\,, \label{n3F} \ee
and it is again covariantly constant because $2\pd_aK=\Gamma^a_{aa}$.

These results should be contrasted with the generic situation (multiple nonvanishing Yukawa couplings) at large complex structure. Generically, the metric is not diagonal, and although the $y_{abc}$ are constant and $D_ay_{bcd}\sim 1/\Im\,t\goesto 0$, the rescaled orthonormal-frame Yukawa couplings are neither constant nor covariantly constant, obeying $\DFab\sim O(1)$. As discussed in Section \ref{sec:D&D}, it is the latter quantities which are actually relevant in moduli stabilization.

\subsection{Flux-modulus variables in terms of fluxes and moduli}
\label{sec:maps}

Using the geometry of the single-Yukawa manifolds described above, we can explicitly construct our models of moduli stabilization. In this subsection, we will start by justifying the claim that any abstract flux-modulus vacuum can be created at any point in the large-complex-structure moduli space of a single-Yukawa manifold, just by tuning fluxes.

Recall that a stabilized vacuum is obtained whenever (the period integrals of) fluxes and moduli $(\tau,t^a)$ are such that the corresponding flux-modulus variables belong to the abstract solution variety ${\cal X}_n$,
\be \big(X(flux,\tau,t^a),\,\Y{i}(flux,\tau,t^a),\,\Z{a}(flux,\tau,t^a)\big) \in {\cal X}_n\,. \ee
In the single-Yukawa case, at large complex structure, we need not worry about matching Yukawa couplings since the orthonormal-frame couplings assume universal, constant values, as in \eqref{n1F}, \eqref{n2F}, \eqref{n3F}. It is entirely reasonable, then, to think that any vacuum $(X,\Y{i},\Z{a})$ can be reached at any fixed values of $(\tau,t^a)$ by adjusting the fluxes, because there are $2n+2$ complex flux-modulus variables and just as many independent complex flux integrals. To show that this is indeed the case, we can consider the form of the actual maps from fluxes and moduli to flux-modulus variables.

For a given Calabi-Yau $Y$, we define the period integrals of the complexified 3-form flux to be
\begin{subequations} \label{comp_flux}
\begin{align}
M^i &= \int_{A_i}G_3 \equiv (m_{RR}^i-\tau m_{NSNS}^i)\,, \\
E_i &= \int_{B^i}G_3 \equiv (e^{RR}_i-\tau e^{NSNS}_i)\,,
\end{align}
\end{subequations}
where $m_{RR}^i,\,e^{RR}_i,\,m_{NSNS}^i,$ and $e^{NSNS}_i$ are integrals of the real $F_{RR}$ and $H_{NSNS}$. The quickest and most general way to obtain the desired maps is to use the original implicit definition of flux-modulus variables via a Hodge decomposition of $G_3$ from \cite{D&DI}, \emph{viz} (in our notation)
\be G_3 \equiv -ie^{-K/2+K_\mbc}\big[X \bar{\Omega} -Y^{\hat{\bar{a}}}\bar{D}_{\hat{\bar{a}}}\bar{\Omega} + \bar{Z}^{\hat{a}}D_{\hat{a}}\Omega - \bar{Y}^{\hat{0}}\Omega\big].
\label{DDG3} \ee
By integrating both sides of \eqref{DDG3} over the $A$- and $B$-cycles of $Y$, we obtain a relation of the form
\be {\scriptsize
\left(\begin{array}{c} M^0 \\ M^1 \\ \vdots \\ E_0 \\ E_1 \\ \vdots \end{array} \right) }
 = -ie^{-K/2+K_\mbc}\,{\cal T}\,\cdot 
 {\scriptsize \left(\begin{array}{c} X \\ \Y{1} \\ \vdots \\
  \cY{0} \\ \cZ{1} \\ \vdots \end{array} \right) },
\label{gen_map}
\ee
where ${\cal T}$ is a matrix that can depend on Yukawa couplings, their covariant derivatives, and other geometric data. Generically, \eqref{gen_map} tells us how to choose fluxes in order to reach any $(\XYZ)$ at any point $(\tau,t^a)$ in moduli space. For our single-Yukawa models, in which points $(\XYZ)\in{\cal X}_n$ describe vacua independently of Yukawa couplings, \eqref{gen_map} then provides the prescription for tuning fluxes to realize any abstract vacuum at any $(\tau,t^a)$ near large complex structure. This justifies our initial claim.

In our models, the above matrix ${\cal T}$ has a fairly simple dependence on complex-structure moduli and the constants $(y_{abc},\,q_{ab},\,\ell_a,\,c)$ appearing in the prepotential \eqref{prep} of $Y$. Its exact form can be constructed from the periods of $\Omega$ and the covariant derivatives on moduli space described in Section \ref{sec:family}. In order to derive the map \eqref{gen_map} more explicitly, however, we find it useful to present a complementary approach starting directly from our definition of flux-modulus variables, \eqref{XYZ}.

As such, first consider the superpotential itself, which is related to $X$ via a rescaling, $X = e^{K/2}W$. Using the expressions for the periods of $\Omega$ near large complex structure given in \eqref{periodsF}, we find that
\begin{align} W &= \int G_3\wedge\Omega = E_iw^i-M^iF_i \nno \\
&= -\frac{1}{6}M^0y_{abc}t^at^bt^c+\frac{1}{2}y_{abc}M^at^bt^c+(q_{ab}M^a+\ell_bM^0+E_b)t^b+(\ell_aM^a+2cM^0+E_0)\,. \label{large_W}
\end{align}
Note that this is a polynomial in the $t^a$, which in the case of a single nonvanishing $y_{abc}$ will always have $1+n+n+1 = 2n+2$ distinct terms. The $2n+2$ coefficients of these terms can always be written as a nonsingular linear combination of the $2n+2$ flux integrals. For example, in the case $n=1$, we can write
\be W = At^3+Bt^2+Ct+D\,, \ee
with
\be \left(\begin{array}{c} A \\ B \\ C \\ D \end{array}\right) =
 \left(\begin{array}{cccc}
  -\frac{1}{6}y_{111} & 0 & 0 & 0 \\
  0 & \frac{1}{2}y_{111} & 0 & 0 \\
  \ell_1 & q_{11} & 0 & 1 \\
  2c & \ell_1 & 1 & 0 \end{array}\right)
  \left(\begin{array}{c} M^0 \\ M^1 \\ E_0 \\ E_1 \end{array}\right).
  \label{n1AM}
\ee
The matrix here has determinant $\sim y_{111}$ and is easily inverted.

The expression for $W$ itself provides the map $X(flux,\tau,t^a)$. To obtain the remainder of the flux-modulus variables $\Y{i} = e^{K/2}\D{i}W$ and $\Z{a} = e^{K/2}\D{a}W$, we can work out the action of covariant derivatives on $W$ using the geometrical data in Section \ref{sec:family}. Generically, the orthonormal-frame derivatives in complex-structure-modulus directions look like%
\footnote{We ignore $\epsilon$-corrections to the covariant derivatives (\emph{cf.} \eqref{def_te}, \eqref{D1111}). These slightly alter some of the maps described here, and should be taken into account if one is interested in computing a specific compactification. Our main goal, however, is simply to illustrate how fluxes can be adjusted to tune flux-modulus variables.}
\be \D{a}W = \Big[\#-\frac{1}{\#}(t^a-\bar{t}^a)\pd_a\Big]W \qquad\mbox{(no sum over $a$)}\,, \ee
turning $W$ into some polynomial involving $t^a$'s and $\bar{t}^a$'s. Similarly, the covariant derivative in the axion-dilaton direction can be calculated from the K\"ahler potential $K_\tau = -\log(-i(\tau-\bar{\tau}))$ to be
\be \D{0}W = [1-(\tau-\bar{\tau})\pd_0]W\,, \ee
which simply acts on the flux-coefficients $A,\,B,\,C,...$, via complex conjugation (there is no $\tau$-dependence elsewhere in $W$). Putting this together, we can write all $2n+2$ flux-modulus variables as linear combinations of the $2n+2$ coefficients $A,\,B,\,C,\,$..., using a linear transformation that depends on $(t^a,\bar{t}^a)$. For example, when $n=1$, we have
\be \left(\begin{array}{c} X \\ \sqrt{3}Y_{\hat{1}} \\ \bar{Y}_{\hat{\bar{0}}} \\ \sqrt{3}\bar{Z}_{\hat{\bar{1}}} \end{array}\right) =
 e^{K/2}\left(\begin{array}{cccc}
 t^3 \,&\, t^2 \,&\, t \,&\, 1 \\
 3t^2\bar{t} \,&\, t^2+2|t|^2 \,&\, 2t+\bar{t} \,&\, 3 \\
 \bar{t}^3 \,&\, \bar{t}^2 \,&\, \bar{t} \,&\, 1 \\
 3\bar{t}^2t \,&\, \bar{t}^2+2|t|^2 \,&\, 2\bar{t}+t \,&\, 3 \end{array}\right)
\left(\begin{array}{c} A \\ B \\ C \\ D \end{array}\right). \label{n1XA}
\ee

This second matrix is also generically nonsingular and easily invertible in every case $n=1,2,3$. By combining the transformations in \eqref{n1AM} and \eqref{n1XA} (or the corresponding expressions for $n=2,3$), we obtain explicit maps from fluxes and moduli to flux-modulus variables. For any fixed desired values of moduli $(\tau,t^a)$, it is straightforward to invert the two matrices and obtain the map from flux-modulus variables to fluxes --- \ie\ the matrix ${\cal T}$ in \eqref{gen_map}. Further details of the case $n=1$, as well as the maps and inverted maps for $n=2$ and $n=3$, can be found in Appendix \ref{app:mapdata}.

In the preceding analysis, we have mostly overlooked the quantization of fluxes. As argued in (\eg) \cite{D&DI}, when the upper bound on the flux-induced tadpole contribution $L$ is large we do not expect quantization to have a great effect. Practically, if we allow large values of the (quantized) real fluxes $m_{RR}^i,\,m_{NSNS}^i,\,e^{RR}_i,$ and $e^{NSNS}_i$, and also allow some freedom in the choice of moduli $(\tau,t^a)$, we should be able to fine-tune at least part of the flux-modulus variables. The real fluxes $m_{RR}^i,\,m_{NSNS}^i,\,e^{RR}_i,$ and $e^{NSNS}_i$ are traditionally quantized as integers, though they may obey a more general quantization if we work with A- and B-cycles forming a non-integral basis of $H_3(Y;\mbr)$ (rather than a basis of $H_3(Y;\mbz)$) in order to make the single-Yukawa condition manifest.

\subsection{Classification of vacua}
\label{sec:X}

Having shown how any values of flux-modulus variables can (in principle) be realized in single-Yukawa flux compactifications at large complex structure, we finally describe the sets of possible vacua in these models --- \ie\ the abstract varieties ${\cal X}_n$ corresponding to solutions of $dV=0$ in flux-modulus variables. This is potentially the most interesting part of the tree-level analysis, since physical properties of vacua are directly linked to the flux-modulus description, as explained in Section \ref{sec:phys}.

Recall from \eqref{derV0} that the critical-point equations are
\bse \label{DV2}
\begin{align}
\pd_{\hat{0}} V = D_{\hat{0}} V &= Z_{\hat{c}} \bar{Y}^{\hat{c}} + Y_{\hat{0}}\bar{X} =0 \,, \\
\pd_{\hat{a}} V = D_{\hat{a}} V &= Z_{\hat{a}}\bar{Y}_{\hat{0}} + \cF_{\hat{a}\hat{b}\hat{c}}\bar{Z}^{\hat{b}}\bar{Y}^{\hat{c}}+Y_{\hat{a}}\bar{X} = 0 \qquad \forall\,\hat{a}\,.
\end{align}
\ese
We have already stressed that, since the orthonormal-frame Yukawa couplings assume universal, constant values at large complex structure, the set of solutions to (\ref{DV2}a-b) can be described entirely in terms of $(X,\Y{i},\Z{a})$ in each case $n=1,2,3$. Thus, the varieties ${\cal X}_n=\{dV=0\}$ become $(2n+2)$-real-dimensional subsets of $(4n+4)$-real-dimensional $(X,\Y{i},\Z{a})$-space.

The equations \eqref{DV2} have the special property that every term is bilinear, involving one of the $2n+2$ variables $\{X,\Z{a},\cX,\cZ{a}\}$, and one of the $2n+2$ variables $\{\Y{i},\cY{i}\}$. This implies that $\Y{i}=0$ $(\forall\,i)$ is always a solution, corresponding to a supersymmetric vacuum (since $\Y{i}\sim \D{i}W$). Likewise, there also always exists a solution $X=\Z{a}=0$ $(\forall\,a)$, which we call ``antisupersymmetric,'' following \cite{D&DI}. It turns out that the antisupersymmetric solution is never physically reasonable due to an inconsistently large cosmological constant (background energy), but we will wait to discuss this until Section \ref{sec:modelanal}.

In addition to the supersymmetric and antisupersymmetric solutions, there also exist ``intermediate'' branches of supersymmetry-breaking solutions of \eqref{DV2}. These solutions can always be parametrized by $n+1$ free phases and $n+1$ free magnitudes (of flux-modulus variables), and are characterized by how the free magnitudes are distributed among the sets $\{X,\Z{a}\}$ and $\{\Y{i}\}$. Equivalently, the solutions are characterized by various relations among the $\{X,\Z{a}\}$ and among the $\{\Y{i}\}$. We will show in Section \ref{sec:modelanal} that these intermediate solutions, particularly those allowing the most freedom among the variables $\{X,\Z{a}\}$ (and imposing the most conditions among the $\{\Y{i}\}$), are the best candidates for physically-reasonable, controllable tree-level vacua.

In the remainder of this subsection, we explicitly display solutions for vacua of our models in the cases $n=1$ and $n=2$, along with expressions for the eigenvalues of the corresponding tree-level moduli mass matrices. Although we have shown in Section \ref{sec:phys} that the eigenvalues must all be positive, since $D\cF=0$, it is useful to have their explicit values for the analysis beyond tree level in Section \ref{sec:analysis}. The case $n=3$ does not present any additional interesting features, but can also be treated explicitly, and appears in Appendix \ref{app:n3sol}.

\subsubsection{$n=1$}

For a single complex modulus, we found in Section \ref{sec:family} that the orthonormal-frame Yukawa coupling at large complex structure is
\be \cF_{\hat{1}\hat{1}\hat{1}} = 2/\sqrt{3}\equiv\cF\,. \ee
The critical-point equations \eqref{DV2} then reduce to%
\footnote{This system is very similar to the one discussed in Section 4.2 of \cite{D&DI}. The main difference is due to our use of the no-scale tree-level potential. We will remark on this further in Section \ref{sec:scale-noscale}.}
\bse \label{n1eqs}
\begin{align}
D_{\hat{0}} V &= \Z1\cY1 +\Y0\cX = 0 \,, \\
D_{\hat{1}} V &= \Z1\cY0+\cF\cZ1\cY1+\Y1\cX = 0\,.
\end{align}
\ese
Straightforward algebra shows that there are four distinct branches of solutions, which we collect in Table \ref{tab:n1sol}, each parametrized by two real magnitudes $(\xi,\nu,...)$ and two real phases $(\alpha,\beta,...)$.

\begin{table}[hbt]
\centering
$\begin{array}{ccccc}
\mbox{Branch} & X & \Y1 & \Y0 & \Z1 
   \vspace{.05cm}\\ \hline  \vspace{-.4cm}\\
\mb{S} & \xi e^{i\alpha} & 0 & 0 & \zeta e^{i\beta} \vspace{.15cm} \\
\mb{\ol{S}} & 0 & \upsilon_1e^{i\beta} & \upsilon_2 e^{i\gamma} & 0 \vspace{.15cm} \\
\mb{A},\mb{A'} & \xi e^{i\alpha} & \upsilon e^{i\beta} & -\lambda_\pm \upsilon e^{i(2\alpha-3\beta)} & \lambda_\pm \xi e^{i(\alpha-2\beta)}
\end{array}  $
\caption{Solutions to $dV=0$ for $n=1$}
\label{tab:n1sol}
\end{table}

\noindent %
The constants $\lambda_\pm$ are defined as
\be \lambda_{\pm} = \frac{1}{2}(|\cF|\pm\sqrt{4+|\cF|^2})=\pm\sqrt{3}^{\pm1}\,. \label{lambda} \ee
Branch $\mb{S}$ is supersymmetric, while branch $\mb{\ol{S}}$ is antisupersymmetric and branches $\mb{A}$ and $\mb{A'}$ are ``intermediate''. 

The eigenvalues of the orthonormal-frame mass matrix $M_{\tau,\mbc}^2$ (\emph{cf.} \eqref{M-decomp}), can be written for the four solutions as
\bse \label{n1evals}
\begin{align} \mb{S}:& \quad (\xi\pm\sqrt{3}\zeta)^2,\,(\xi\pm\frac{1}{\sqrt{3}}\zeta)^2 \label{S1} \\
\mb{\ol{S}}: &\quad(\upsilon_0\pm\sqrt{3}\upsilon_1)^2,\,(\upsilon_0\pm\frac{1}{\sqrt{3}}\upsilon_1)^2 \\
\mb{A}: &\quad 16\xi^2,\, 4(\xi^2+3\upsilon^2),\,\frac{16}{3}\upsilon^2,\,\frac{4}{3}(\upsilon^2+3\xi^2) \\
\mb{A'}: &\quad \frac{16}{9}\xi^2,\, \frac{4}{9}(\xi^2+3\upsilon^2),\,\frac{16}{3}\upsilon^2,\,\frac{4}{3}(\upsilon^2+3\xi^2)\,.
\end{align}
\ese
We see explicitly that all eigenvalues are positive, and that all tree-level vacua are (meta)stable.

\subsubsection{$n=2$}

For $n=2$ and orthonormal-frame Yukawa coupling
\be \cF_{\hat{1}\hat{1}\hat{2}}=1\,, \ee
the critical-point equations are
\begin{subequations}
\begin{align} D_{\hat{0}}V &= \Z1\cY1+\Z2\cY2+\Y0\cX = 0 \,, \\
 D_{\hat{1}}V &= \Z1\cY0+\cY1\cZ2+\cY2\cZ1+\Y1\cX = 0 \,, \\
 D_{\hat{2}}V &= \Z2\cY0+\cY1\cZ1+\Y2\cX = 0 \,.
\end{align} \label{n2eqs}
\end{subequations}
These now have eight distinct branches of solutions. The easiest way to find them is to write each flux-modulus variable in terms of a magnitude and a phase, require for each equation that every term has the same phase (\ie\ that the phases factor out), and then solve for the phases and magnitudes separately. We arrive in this way at the parametrizations in Table \ref{tab:n2sol}.

\begin{table}[b]
\centering
$\begin{array}{ccccccc}
\mbox{Branch} & X & \Y1 & \Y2 & \Y0 & \Z1 & \Z2 
   \vspace{.05cm}\\ \hline  \vspace{-.4cm}\\
\mb{S} & \xi e^{i\alpha} & 0 & 0 & 0 & \zeta_1 e^{i\gamma} & \zeta_2 e^{i\beta} \vspace{.1cm} \\
\mb{\ol{S}} & 0 & \upsilon_1 e^{i\alpha} & \upsilon_2 e^{i\beta} & \upsilon_0 e^{i\gamma} & 0 & 0 \vspace{.1cm} \\
\mb{A},\,\mb{A}' & \xi e^{2i\alpha} & \mp\sqrt{2}\upsilon e^{i(\alpha-\gamma)} & \ds\upsilon e^{i\beta} & \ds\upsilon e^{i(2\alpha-\beta+2\gamma)} & \ds\pm\frac{\xi+\zeta}{\sqrt{2}}e^{i(\alpha-\beta+\gamma)} & \zeta e^{2i\gamma} \vspace{.1cm} \\
\mb{B},\,\mb{B}' & \xi e^{2i\alpha} & \ds\pm\frac{\upsilon_0+\upsilon_2}{\sqrt{2}} e^{i(\alpha-\gamma)} & \upsilon_2 e^{i\beta} & \upsilon_0 e^{i(2\alpha-\beta+2\gamma)} & \mp\sqrt{2}\xi e^{i(\alpha-\beta+\gamma)} & \xi e^{i\gamma} \vspace{.1cm} \\
\mb{C} & \xi e^{2i\alpha} & \upsilon_1 e^{i(\alpha-\gamma)} & \upsilon_2 e^{i\beta} & \upsilon_2 e^{i(2\alpha-\beta+2\gamma)} & 0 & -\xi e^{i\gamma} \vspace{.1cm} \\
\mb{D} & \xi e^{2i\alpha} & 0 & \upsilon e^{i\beta} & -\upsilon e^{i(2\alpha-\beta+2\gamma)} & \zeta e^{i(\alpha-\beta+\gamma)} & \xi e^{i\gamma}
\end{array} $
\caption{Solutions to $dV=0$ for $n=2$}
\label{tab:n2sol}
\end{table}
%

Again, there is a supersymmetric branch $\mb{S}$ and an antisupersymmetric branch $\mb{\ol{S}}$, in addition to six intermediate supersymmetry-breaking branches. Branches $\mb{A}$, $\mb{A'}$, and $\mb{D}$ have the most freedom among the variables $X$, $\Z{1}$, and $\Z{2}$.
The eigenvalues of the orthonormal-frame mass matrix $M_{\tau,\mbc}^2$ are
\bse \label{n2evals}
\begin{align}
\mb{S}: &\quad (\xi\pm\zeta_2)^2,\,\,(\xi\pm\sqrt{2}\zeta_1\pm\zeta_2)^2,\,\,(\xi\pm\sqrt{2}\zeta_1\mp\zeta_2)^2 \vspace{.2cm} \\
\mb{\ol{S}}: &\quad (\upsilon_0\pm\upsilon_2)^2,\,\,(\upsilon_0\pm\sqrt{2}\upsilon_1\pm\upsilon_2)^2,\,\,(\upsilon_0\pm\sqrt{2}\upsilon_1\mp\upsilon_2)^2 \vspace{.25cm} \\
\mb{A},\mb{A}': &\quad (\xi+\zeta)^2+4\upsilon^2,\,\, 4(\xi+\zeta)^2,\,\, (\xi-\zeta)^2,\,\, 4(\xi^2+\upsilon^2),\,\, 4(\zeta^2+\upsilon^2),\,\, 16\upsilon^2 \vspace{.25cm} \\
\mb{B},\mb{B}': &\quad (\upsilon+\upsilon_2)^2+4\xi^2,\,\, 4(\upsilon+\upsilon_2)^2,\,\, (\upsilon-\upsilon_2)^2,\,\, 4(\upsilon^2+\xi^2),\,\, 4(\upsilon_2^2+\xi^2),\,\, 16\xi^2 \vspace{.25cm} \\
\mb{C}: &\quad 4\xi^2,\,\,2\upsilon_1^2,\,\,2(2\xi^2+\upsilon_1^2),\,\,4(\xi^2+\upsilon_2^2),\,\,2(\upsilon_1\pm\sqrt{2}\upsilon_2)^2 \vspace{.25cm}\\ 
\mb{D}: &\quad 2\zeta^2,\,\, 4\upsilon^2,\,\, 2(\zeta\pm\sqrt{2}\xi)^2,\,\,4(\xi^2+\upsilon^2),\, 2(\zeta^2+2\upsilon^2) \,.
\end{align}
\ese
As expected, they are all explicitly positive.

\section{Putting tree-level models in perspective}
\label{sec:analysis}

The tree-level models just described are computationally appealing, but they are incomplete. Here, we want to focus on the fact that they do not include potentially significant stringy corrections, which are needed to stabilize K\"ahler moduli.

This is a general problem of type IIB compactifications (see \eg\  \cite{Denef}). The only dependence on K\"ahler moduli in the tree-level no-scale potential \eqref{V-noscale} comes from the prefactor $e^{K_K}\sim 1/{\cal V}^2$. Thus, a tree-level vacuum that stabilizes complex-structure and axion-dilaton moduli supersymmetrically ($\D{i}W =\Y{i}=V=0$) has a flat potential for K\"ahler moduli; whereas a nonsupersymmetric tree-level vacuum ($\D{i}W,\,\Y{i},\,V\neq 0$) always appears to run to infinite volume, \ie\ to decompactify. Neither situation is acceptable. Fortunately, the no-scale structure is generically broken by stringy corrections, which depend on $\alpha'$, and these corrections can be controlled as long as they actually stabilize the internal volume at a large value
\be \langle \mbox{Vol}(Y)\rangle = R^6 \gg \alpha'^3\,. \label{largeV2} \ee
We recall from Section \ref{sec:conv} that \eqref{largeV2} is also necessary to put the Kaluza-Klein compactification scale below the string scale. A complete, consistent compactification of type IIB string theory must look beyond tree level (defined as $O((\alpha'/R^2)^0)$), include $\alpha'$ corrections to the scalar potential, and realize \eqref{largeV2}.

Several fruitful studies of K\"ahler stabilization beyond tree level have been conducted and refined in recent years. In particular, \cite{BB, BBCQ} initiated a program of K\"ahler stabilization at large volume using the explicit form of leading $\alpha'$ corrections. These studies typically assume complex-structure and axion-dilaton moduli to be fixed supersymmetrically at tree level, and then use these moduli, along with vacuum value of the tree-level superpotential, as fixed, tunable parameters. In this section, we conduct a much simpler but also more general analysis of corrections to the tree-level scalar potential. Our emphasis is not so much on obtaining explicit large-volume K\"ahler vacua, but on how $\alpha'$ corrections affect the initial \emph{tree-level} structure. We ask whether it is \emph{possible} for various tree-level vacua to form meaningful (\ie\ controllable) and consistent foundations for more complete stabilizations beyond tree level.

The main analysis is carried out in Section \ref{sec:approx}, and its results are summarized in Section \ref{sec:summ}. In Section \ref{sec:modelanal}, we then apply these results to our large-complex-structure models in order to properly evaluate their potential usefulness and physical properties. We will continue using the tree-level flux-modulus notation of previous sections throughout.

\subsection{Effects of correcting tree-level structure}
\label{sec:approx}

We begin by defining two constants $\delta$ and $\eta$ which capture the rough order of magnitude of corrections to the tree-level K\"ahler potential and superpotential, respectively, in type IIB orientifold compactifications. The ${\cal N}=1$ K\"ahler potential receives perturbative stringy corrections which are suppressed at large volume by powers of the dimensionless ratio $\alpha'/R^2$ \cite{Becker:2002nn, Berg:2005ja, Berg:2005yu, BHP, CCQ}.%
\footnote{The perturbative corrections may be expanded as a series in both $\alpha'/R^2$ and $g_s$. However, as noted in Section \ref{sec:conv}, $g_s$ corrections (from string loops) are believed to always be accompanied by two or more powers of $\alpha'/R^2$ as well \cite{Berg:2005ja, BHP, CCQ}. For this analysis, we can ignore the ``subdominant'' $g_s$ expansion.} %
In particular, leading corrections are at most of order $\alpha'^2/R^4$, so we define
\be \delta \equiv \frac{\alpha'^2}{R^4} \ll 1 \ee
and write the full K\"ahler potential as
\be \framebox{$ K = K_0 + K_p\,, \qquad K_p = O(\delta)\,, $} \label{K} \ee
where $K_0$ denotes the tree-level part as in \eqref{K0}. The ${\cal N}=1$ superpotential is also corrected, but only nonperturbatively \cite{Burgess:2005jx} by effects such as Euclidean D3-instantons \cite{Berglund:2005dm} or gaugino condensation on D7 branes \cite{JL, Gorlich:2004qm}. Regardless of their origin, the corrections to $W$ must be suppressed by powers of\, $\exp\hspace{-.08cm}\left[-\mbox{Vol}(\Sigma_4)/\alpha'^2\right]$, for various 4-cycles $\Sigma_4$ in $Y$. Therefore, we will write
\be \framebox{$ {W} = {W}_0 + {W}_{np}\,,\qquad \tW_{np}= O(\eta)\,, $} \label{W} \ee
where $W_0$ is the tree-level superpotential \eqref{W0}, and we expect that
\be \eta \sim e^{-1/\delta} \lll  \delta\,, \ee
as long as $Y$ is not too anisotropic. The tilde in \eqref{W} indicates rescaling by $e^{K/2}$ as in previous sections, \ie\ $\tW_{np}=e^{K/2}W_{np}$. It is convenient to define $\eta$ this way since we will always be comparing it to other rescaled quantities.%
\footnote{Recall that there is no large-volume factor $\alpha'^6/\langle\mbox{Vol}(Y)\rangle^2 \sim \delta^{3}$ coming from the K\"ahler-moduli piece $e^{K_K}$ of rescalings by $e^K$ (or $e^{K_0}$). As explained in Section \ref{sec:conv}, this factor has been explicitly removed in our conventions, so that $e^{K_K}$ is roughly $O(1)$.}

Our plan now is to examine how the corrections in \eqref{K} and \eqref{W} propagate to important quantities such as the scalar potential and its derivatives, extending a similar treatment in \cite{CQS}. Specifically, we define flux-modulus variables at \emph{tree level} via
\bse
\begin{align}
X &= e^{K_0/2}W_0 \,, \\
Y_{{i}} &= e^{K_0/2}D_{{i}}^{(0)}W_0 \,, \\
Z_{{a}} &= e^{K_0/2}D_{{0}}^{(0)}D_{{a}}^{(0)}W_0 \,,
\end{align}
\ese
consistent with their use in Sections \ref{sec:formalism} and \ref{sec:MODELS}, and we seek to write $D_AW,\,V,\,D_AV$, etc. in terms of these variables plus leading corrections of order $\delta$ and $\eta$.

In an orthonormal frame, the $\delta$-corrections to $\D{A}\tW,\,V,\,\D{A}V$, etc. come entirely from covariant derivatives and rescaling factors $e^K$ or $e^{K/2}$. The latter contributions are almost trivial, since $e^K,\,e^{K/2}$ commute with covariant derivatives, and can be factored out of all important quantities. At leading order, we simply have
\be e^K \sim (1+\delta)e^{K_0}\,.
\ee

As for covariant derivatives, they can schematically be expanded as
\be
D_{\hat{A}} = D_{\hat{A}}^{(0)} + (\delta) \,c{_{\hat{A}}}^{\hat{B}}D_{\hat{B}}^{(0)} + (\delta)\,c'{_{\hat{A}}}\,,
\label{D-corr}
\ee
where the $O(1)$ tensors $c$ and $c'$ contain the combined corrections to the vielbein and the K\"ahler and metric connections. The action of \eqref{D-corr} on the tree-level $W_0$ is fairly straightforward, as we shall see below, but the action on $W_{np}$ merits some comments. To approximate $D_{\hat{A}}\tW_{np}$, only the zeroth-order piece $D_{\hat{A}}^{(0)}$ is necessary, since $\tW_{np}$ is already $O(\eta)$. In Sections \ref{sec:family} and \ref{sec:maps}, we saw that the orthonormal-frame covariant derivative in complex-structure and axion-dilaton directions typically looks like
\be D_{\hat{i}}^{(0)} \sim a+b\,(\Im\,t^i)\,\pd_i\,, \qquad a,\,b \sim O(1)\,, \ee
so that $D_{\hat{i}}^{(0)}$ itself is $O(1)$ when acting on any algebraic functions of $(\tau,t^a)$, even near large complex structure or weak coupling. Then, assuming $\tW_{np}$ is indeed algebraic in $(\tau,t^a)$, we have at leading order
\be \D{i}\tW_{np} \sim \eta\,. \ee
The K\"ahler potential for K\"ahler moduli (near large volume) has the same structure as the K\"ahler potential for complex moduli (near large complex structure), namely $K_K \sim -\log[(\Im\,\rho^\alpha)^3]$, so we also expect that
\be D_{\hat{\alpha}}^{(0)} \sim a' + b'\,(\Im\,\rho^\alpha)\,\pd_\alpha\,, \qquad a',\,b'\sim O(1)\,, \ee
with $D_{\hat{\alpha}}^{(0)}$ just being $O(1)$ when acting on algebraic functions of the $\rho^\alpha$, even near large volume. However, $W_{np}$ involves the K\"ahler moduli \emph{exponentially}, so in fact
\be \D{\alpha}\tW_{np} \sim (\Im\,\rho^\alpha)\tW_{np} \sim \delta^{-1}\eta\,, \ee
in contrast (for example) to
\be D_{\hat{\alpha}}^{(0)} W_0 \sim W_0\,. \ee
These approximations should hold around large complex structure and large volume, or more generally as long as we stay away from any singularities such as conifold points in the Calabi-Yau moduli space.

Putting all this together, we can approximate the leading corrections to $\D{A}\tW,\,V,\,\D{A}V$, etc. by just substituting
\begin{align}
\tW &\mapsto \tW_0 + O(\eta)\,, \\
\D{A} &\mapsto D_{\hat{A}}^{(0)}+O(\delta)\,D_{*}^{(0)}+O(\delta)\,,
\end{align}
and being careful about K\"ahler-derivatives of ``$\eta$''. We find, for example, that
\bse \label{Wcorr}
\begin{align}
\tW &= X + O(\eta)\,, \label{Xcorr} \\
\D{i} \tW &= \Y{i} + \delta\,c{_{\hat{i}}}^{\hat{j}}D_{\hat{j}}^{(0)}\tW_0 + \delta\, c'_{\hat{i}}\tW_0 + \D{i}\,\eta \nno \\
  &= \Y{i} + O(\delta|Y|+\delta|X|+\eta)\,, \\
\D0 \D{a}\tW &= \Z{a}+O(\delta|Z|+\delta|Y|+\delta^2|X|+\eta)\,, \label{ferm1} \\
\D{a}\D{b}\tW &= \cF_{\hat{a}\hat{b}\hat{c}}\bar{Z}^{\hat{c}}+O(\delta|Z|+\delta|Y|+\delta^2|X|+\eta)\,, \label{ferm2} \\
\D{\alpha} \tW &= D_{\hat{\alpha}}^{(0)}\tW_0 + \delta\,c{_{\hat{\alpha}}}^{\hat{\beta}}D_{\hat{\beta}}^{(0)}\tW_0 + \delta\, c'_{\hat{\alpha}}\tW_0 + \D{\alpha}\,\eta \nno \\
  &= O(1)\cdot X + O(\delta|X|+\delta^{-1}\eta) \nno \\ &\sim X+\delta|X|+\delta^{-1}\eta\,, \label{DaWcorr}
\end{align}
\ese
with $|Y|=(|Y_{\hat{i}}|^2)^{1/2}$ and $|Z|=(|Z_{\hat{a}}|^2)^{1/2}$ denoting typical magnitudes of the $\Y{i}$ and $\Z{a}$, respectively.%
\footnote{For very large numbers of complex structure moduli, one should be careful about extra numerical factors $\sim n$ entering these equations as well.} %
Continuing this process with higher derivatives of $\tW$ and substituting the answers into the general expressions for $V$ and its derivatives from \eqref{DVgen}, we obtain the more interesting
\bse \label{Vcorr}
\begin{align}
V &= |Y|^2 + O\big(\,\delta|X|^2+\delta|Y|^2+\delta|X||Y|+\delta^{-1}\eta|X|+\eta|Y| + \delta^{-2}\eta^2 
  \,\big) \,, \label{corr_V} \\
D_{\hat{i}}V &= D^{(0)}_{\hat{i}}V_0 + O\big(\, \delta|X|^2+\delta|Y|^2+\delta|X||Y|+\delta|Y||Z|+\delta|X||Z| \label{corr_DCV} \\
 &\hspace{6.5cm} +\delta^{-1}\eta|X|+\delta^{-1}\eta|Y|+\eta|Z| +\delta^{-2}\eta^2 
  \,\big) \,, \nno \\
D_{\hat{\alpha}}V &= (\pd_{\hat{\alpha}}K_0)|Y|^2 + O\big(\, \delta|X|^2+\delta|Y|^2+\delta|X||Y|+\delta^{-2}\eta|X|+\delta^{-1}\eta|Y|+\delta^{-3}\eta^2 
  \,\big)\,. \label{corr_DKV} 
\end{align}
\ese
The form of \eqref{corr_DCV} implies, in particular, that the values of the flux-modulus variables in a full vacuum are corrected from their values in a tree-level vacuum roughly up to the scale
\be {\cal C} = \delta|X|+\delta|Y|+\delta|Z|+\delta^{-1}\eta \,. \label{C} \ee
This tells us approximately how much control we have over tree-level structure.

We will now proceed to use the order-of-magnitude approximations \eqref{Wcorr} and \eqref{Vcorr} to analyze the consistency and physical properties of various complete vacua which could potentially be constructed from tree-level models. Our goal, again, is to use consistency and physical requirements to impose conditions on the tree-level vacua. We begin with supersymmetric vacua in Section \ref{sec:susy-corr}, then look at nonsupersymmetric vacua in Section \ref{sec:nonsusy-corr}, and finally try to extract some conditions for stability in Section \ref{sec:stab}.

\subsubsection{Supersymmetric vacua}
\label{sec:susy-corr}

Let us first consider the simplest case of supersymmetric vacua. Suppose we start with a supersymmetric tree-level vacuum in the axion-dilaton/complex-structure directions,
\be |Y|_{vac} = 0 \,, \ee
which we want to correct to a fully stabilized vacuum satisfying $D_{\hat{\alpha}}W|_{vac} = 0$ as well.%
\footnote{We use ``$|_{vac}$'' throughout Sections \ref{sec:susy-corr} and \ref{sec:nonsusy-corr} as a reminder that equations here only apply to the vacuum values of various quantities, which can sometimes be very nongeneric.} %
From \eqref{DaWcorr}, we see that we need roughly
\be |X|_{vac} + \delta|X|_{vac} + \delta^{-1}\eta \sim 0\,, \label{susycorr} \ee
\ie\ the three terms of orders $|X|_{vac}$, $\delta|X|_{vac}$, and $\delta^{-1}\eta$ must somehow cancel. One possibility is that
\be \framebox{$|X|_{vac} \sim \delta^{-1}\eta $}\,, \label{KKLT} \ee
which is precisely the KKLT scenario: $|X|_{vac}$, in other words $|\tW_0|_{vac}$, is tuned to a parametrically small value. Note, however, that this places $|X|_{vac}$ below the correction scale ${\cal C}$, so its actual value will depend strongly on the details of K\"ahler stabilization. (This potential shortcoming of KKLT scenarios was also noted in \cite{deAlwis:2005tf}). In fact, relation \eqref{susycorr} only needs to hold up to corrections of order ${\cal C}|_{vac}$, which suggests a second possibility. We could also have
\be \framebox{$ |X|_{vac} \lesssim \delta|Z|_{vac} $} \qquad \imp \quad |X|_{vac} \lesssim {\cal C}\,, \label{KKLT2} \ee
effectively ``swamping out'' any constraint coming from \eqref{susycorr}.%
\footnote{It might be interesting to work out a specific example of stabilization corresponding to this second regime as an alternative to KKLT. Simultaneous stabilization of complex-structure and K\"ahler moduli may be necessary.} %

In either case \eqref{KKLT} or \eqref{KKLT2}, the exact solution to $D_{\hat{\alpha}}W=0$ will certainly depend on the specific form of corrections to $K$ and $W$. These two conditions simply allow the \emph{possibility} that a fully-stabilized solution may be constructed.

As for physical requirements, \eqref{corr_V} shows that in a supersymmetric vacuum the cosmological constant is
\be \frac{|\Lambda|}{M_s^4} = \frac{|V|_{vac}}{M_s^4} \sim \delta|X|_{vac}^2+\delta^{-1}\eta|X|_{vac}+\delta^{-2}\eta^2\,, \label{susy_lambda} \ee
again up to ${\cal C}$-corrections. Remember that the dimensionful scalar potential \eqref{V} contains a prefactor $T_3\sim (\alpha')^{-2}\sim M_s^4$; we just suppressed this in (\eg) \eqref{Vcorr}. As long as $|X|_{vac}$ is not too large, \eqref{susy_lambda} implies that $|\Lambda| \ll M_s^4$, which is necessary for consistency of our effective field theory. (Flux quantization typically produces flux-modulus variables whose magnitudes are very roughly $O(1)$, so imposing that $|X|^2$ is much smaller than $\delta^{-1}$ (say) is not unreasonable.) Since both scenarios \eqref{KKLT} and \eqref{KKLT2} have $|X|$ below the correction scale ${\cal C}$, any further tuning of $\Lambda$ to parametrically small values many orders of magnitude below the string scale is controlled by the details of stringy corrections. However, in a fully supersymmetric vacuum, the cosmological constant will always be AdS.

Chiral modulino masses $(\alpha'/R^3)\D{i}\D{j}\tW$ (\emph{cf}. \eqref{modulinos}) have rough orders of magnitude
\be M_{\tilde{\mbc},\tilde{\tau}}\sim \frac{\alpha'}{R^3}(|Z|+\delta^2|X|+\delta|Z|+\eta)\,, \ee
as can be seen from \eqref{ferm1} and \eqref{ferm2}. Requiring them to be below the string, Planck, or Kaluza-Klein scales puts a loose bound on $|Z|$, which can be important when trying to achieve hierarchies like \eqref{KKLT2} in specific compactifications.

\subsubsection{Nonsupersymmetric vacua}
\label{sec:nonsusy-corr}

Turning to nonsupersymmetric vacua, there are two choices: either supersymmetry is broken at tree level ($|Y|_{vac} \neq 0$), or it is preserved at tree level ($|Y|_{vac}= 0$) but broken by the fixing of K\"ahler moduli ($D_{\hat{\alpha}}W|_{vac}\neq0$).

The second situation is similar to the supersymmetric vacua above, but instead of \eqref{susycorr} we must satisfy
\be \pd_{\hat{\alpha}}V = D_{\hat{\alpha}}V \sim \delta|X|^2_{vac}+\delta^{-2}\eta|X|_{vac}+\delta^{-3}\eta^2 \sim 0 \ee
in order to stabilize K\"ahler moduli. Scenarios \eqref{KKLT} and \eqref{KKLT2}, which have $|X|\lesssim {\cal C}$, are still acceptable. However, there now arises another possibility,
\be \framebox{$ |X|_{vac} \sim \delta^{-2}\eta $} \quad \gg\delta^{-1}\eta \,. \label{BBCQ}
\ee
This seems to correspond roughly to the large-volume AdS vacua of \cite{BBCQ}, which are characterized by values of $|W|$ significantly larger than those in KKLT scenarios. If we also have $|Z|_{vac}\lesssim\delta^{-1}|X|_{vac}$ (for example, if $|Z|_{vac}\sim |X|_{vac}$), then we can keep $|X|_{vac}$ above the correction scale ${\cal C}$ and maintain tree-level control over both $X$ and the $\Z{a}$ in a given model.  The cosmological constant here, again given by \eqref{susy_lambda}, should have no problem being below the string scale. A more careful analysis indeed shows that when $|X|>{\cal C}$ it is always negative. Unfortunately, to obtain parametrically small (and possibly dS) values of $\Lambda$ without resorting to a KKLT-like uplifting mechanism, it is necessary to go back to $|X|\lesssim{\cal C}$ as in \eqref{KKLT} or \eqref{KKLT2}. (For a realization of a dS cosmological constant due to F-term supersymmetry-breaking in the K\"ahler sector, see \eg\ \cite{Misra:2007yu}.)

Of more interest to us in this paper are the tree-level supersymmetry-breaking vacua, since most of the branches of our models in Section \ref{sec:X} fall into this category. From \eqref{corr_DKV}, the condition for allowing K\"ahler stabilization when $|Y|_{vac}\neq0$ becomes
\be D_{\hat{\alpha}}V \sim |Y|^2_{vac} + \delta|X|^2_{vac}+ \delta|Y|^2_{vac}+ \delta|X|_{vac}|Y|_{vac}+\delta^{-2}\eta|X|_{vac}+\delta^{-1}\eta|Y|_{vac}+\delta^{-3}\eta^2 \sim 0\,. \label{nosusycond} \ee
The leading term $|Y|^2_{vac}$ must cancel against one of the corrections, leading to the conditions%
\footnote{Another possibility is $|Y|^2_{vac}\lesssim \delta^{-2}\eta|X|_{vac}$, but this is only distinct from \eqref{my1} if $|X|\lesssim \delta^{-1}\eta$, which would then place $|X|_{vac},|Y|_{vac}\lesssim {\cal C}$, resulting in the loss of tree-level control over both $X$ and the $\Y{i}$.}
\be \framebox{$ |Y|_{vac}^2 \,\lesssim\, \delta|X|^2_{vac} $} \label{my1} \ee
or
\be \framebox{$ |Y|_{vac} \,\lesssim\, \delta^{-1/2}\cdot \delta^{-1}\eta $}\,. \label{my2} \ee
Either of these can be satisfied without completely swamping out $|Y|_{vac}$ by corrections; for example, in the case of \eqref{my1}, having $|Y|_{vac}\sim\sqrt{\delta}|X|_{vac}\gg \delta|X|_{vac}$ could keep $|Y|_{vac}\gg {\cal C}$. Having $|Y|_{vac}\lesssim{\cal C}$ is of course possible as well, but then we lose control over the supersymmetry-breaking scale, arguably the most important characteristic of this class of vacua.

The cosmological constant for corrected nonsupersymmetric tree-level vacua is
\be \frac{|\Lambda|}{M_s^4} \sim |Y|_{vac}^2+\delta|X|_{vac}^2+\delta|Y|_{vac}^2+\delta|X|_{vac}|Y|_{vac}+\delta^{-1}\eta|X|_{vac}+\eta|Y|_{vac}+\delta^{-2}\eta^2. \label{ns-Lambda} \ee
This is similar but not identical to the RHS of \eqref{nosusycond}. Unlike the previous cases with tree-level supersymmetry, an additional cancellation is necessary to keep $\Lambda$ below the string scale and consistent with effective field theory. Condition \eqref{my1} is sufficient to allow such a cancellation, \emph{though condition \eqref{my2} is not}. Depending on the precise form of corrections, the cosmological constant can be either dS or AdS.

We can consider the scale of supersymmetry breaking and the gravitino mass for tree-level supersymmetry-breaking solutions as well. From \eqref{susygrav}, replacing $|\Y{i}|^2$ and $|X|^2$ by the actual quantities $|D_{\hat{A}}\tW|^2$ and $|\tW|^2$ (respectively), we find
\begin{align} \frac{M_{susy}^4}{M_P^4} &\sim \delta^3\left[|Y|^2+3|X|^2 + \delta|X|^2+\delta|Y|^2+\delta|X||Y|+\eta|Y|+\delta^{-1}\eta|X|+\delta^{-2}\eta^2\right]_{vac} \label{ns-susy}\,, \\
\frac{M_{3/2}^2}{M_P^2} &\sim \delta^3\left[|X|^2+\eta|X|+\eta^2\right]_{vac}\,. \label{ns-grav}
\end{align}
The leading terms $|Y|^2+3|X|^2$ and $|X|^2$, respectively, are given exactly. Generically, we would expect that $M_{susy}^4/M_P^4 \sim \delta^3$ and $M_{3/2}^2/M_P^2 \sim \delta^3$, which may be acceptable depending on the desired scale of supersymmetry breaking in the visible sector and the mediation mechanism employed. In order to obtain values \emph{many} orders of magnitude smaller than $M_P$, there need to be cancellations between the leading terms and corrections, which forces $|X|_{vac}\lesssim{\cal C}$. Due to \eqref{my1} (since \eqref{my2} is not an option), this implies that $|Y|_{vac}\lesssim{\cal C}$ as well. To achieve $|X|_{vac}\lesssim{\cal C}$, we can either have $|X|_{vac}\lesssim\delta^{-1}\eta$\, or, more likely,
\be \framebox{$ |X|_{vac} \lesssim \delta |Z|_{vac} $}\,. \label{XZswamp} \ee
We then only retain control over the $\Z{a}$.

Akin to the supersymmetric case, chiral modulino masses are of order
\be M_{\tilde{\mbc},\tilde{\tau}}\sim \frac{\alpha'}{R^3}(|Z|+\delta^2|X|+\delta|Y|+\delta|Z|+\eta). \ee
Since physically reasonable tree-level vacua seem to favor a hierarchy of parameters such as $\delta^{-1/2}|Y|\lesssim |X|\lesssim \delta|Z|$, a bound on $|Z|$ coming from $|M_{\tilde{\mbc},\tilde{\tau}}|\lesssim M_s,M_P,M_{KK}$ can be significant.

\subsubsection{Stability}
\label{sec:stab}

Let us finally consider how stringy corrections affect the stability of tree-level vacua. Including K\"ahler moduli, the complete mass matrix in an orthonormal frame is
\be {\textstyle \left(\frac{R^6}{\alpha'^2}\right) } M^2 = \left(\begin{array}{cccc}
 \ol{D}_{\hat{\ol{i}}}D_{\hat{j}}V & \ol{D}_{\hat{\ol{i}}}\ol{D}_{\hat{\ol{j}}} V &  \ol{D}_{\hat{\ol{i}}}D_{\hat{\beta}}V & \ol{D}_{\hat{\ol{i}}}\ol{D}_{\hat{\ol{\beta}}} V \\
 D_{\hat{i}} {D}_{\hat{{j}}} V & D_{\hat{i}}\ol{D}_{\hat{\ol{j}}}V & D_{\hat{i}} {D}_{\hat{{\beta}}} V & D_{\hat{i}}\ol{D}_{\hat{\ol{\beta}}}V \\
\ol{D}_{\hat{\ol{\alpha}}}D_{\hat{j}}V & \ol{D}_{\hat{\ol{\alpha}}}\ol{D}_{\hat{\ol{j}}} V &  \ol{D}_{\hat{\ol{\alpha}}}D_{\hat{\beta}}V & \ol{D}_{\hat{\ol{\alpha}}}\ol{D}_{\hat{\ol{\beta}}} V \\
 D_{\hat{\alpha}} {D}_{\hat{{j}}} V & D_{\hat{\alpha}}\ol{D}_{\hat{\ol{j}}}V & D_{\hat{\alpha}} {D}_{\hat{{\beta}}} V & D_{\hat{\alpha }}\ol{D}_{\hat{\ol{\beta}}}V 
 \end{array}\right)\Big|_{vac} \equiv 
 \left(\begin{array}{cc} M_{\mbc,\tau}^2 & S \\ S^\dagger & M_K^2
 \end{array}\right).
\ee
The matrix $M^2$ is positive-definite only if the diagonal blocks $M^2_{\mbc,\tau}$ and $M^2_K$ are individually positive-definite, and if the off-diagonal blocks $S$ and $S^\dagger$ don't destabilize the eigenvalues of $M^2_{\mbc,\tau}$ and $M^2_K$ too much. At tree level, $M_{\mbc,\tau}^2$ is just the matrix in \eqref{M-decomp} and the off-diaganal part $S$ vanishes, decoupling the axion-dilaton/complex-structure moduli from the K\"ahler moduli. Including the stringy corrections to $K$ and $W$, we find that the elements of $M_{\mbc,\tau}^2$, $S$, and $M_K^2$ generally have magnitudes
\bse
\begin{align} M_{\mbc,\tau}^2 &\sim (M_{\mbc,\tau}^2)_0 + 
 \delta(|X|+|Y|+|Z|)^2+\delta^{-1}\eta(|X|+|Y|+|Z|)+\delta^{-2}\eta^2\,, \label{Mtc-corr} \\
 S &\sim 0+
 \delta(|X|+|Y|+|Z|)^2+\delta^{-2}\eta(|X|+|Y|+|Z|)+\delta^{-3}\eta^2\,,
  \label{Scorr} \\
 M_K^2 &\sim (M_K^2)_0+
 \delta(|X|+|Y|+|Z|)^2+\delta^{-3}\eta(|X|+|Y|+|Z|)+\delta^{-4}\eta^2\,.
\end{align}
\ese
The negative powers of $\delta$ are just determined by the number of covariant K\"ahler derivatives that can act on $W_{np}$'s. Comparing \eqref{Mtc-corr} and \eqref{Scorr}, we therefore estimate 
\be {\cal C}' = \delta(|X|+|Y|+|Z|)^2+\delta^{-2}\eta(|X|+|Y|+|Z|)+\delta^{-3}\eta^2 \ee
to be the scale up to which we can expect eigenvalues of $M^2$ coming from $M_{\mbc,\tau}^2$ to be perturbed.

From this simple analysis, all we can say is that \emph{if} the tree-level eigenvalues of $M^2_{\mbc,\tau}$ are all positive and greater than $\cal C'$, then the corresponding eigenvalues of $M^2$ will remain positive as well. Otherwise, stability of the axion-dilaton/complex-structure moduli will be interdependent with K\"ahler stabilization. Certainly, stability of the K\"ahler moduli themselves will always depend on the details of $K_{p}$ and $W_{np}$.%
\footnote{For some further general constraints on stability, considering in particular vacua with supersymmetry breaking in the K\"ahler sector and the corresponding sGoldstino mass, see \cite{CGGLPS}. Stability in the K\"ahler sector is also treated in \cite{CQS} and related works.} %
On the other hand, if an axion-dilaton/complex-structure vacuum is \emph{not} stable at tree level, it has very little chance of regaining stability after being corrected.

Note that condition \eqref{my1}, which is necessary for tree-level supersymmetry-breaking vacua, implies that $|Y|^2 \lesssim {\cal C}'$. (Any potential alternatives to \eqref{my1} having $|Y|_{} \lesssim {\cal C}$ imply $|Y|_{}^2 \lesssim{\cal C}'$ anyway.) Therefore, no eigenvalue of $(M_{\mbc,\tau}^2)_0$ controlled by $|Y|^2$ can ever be guaranteed stability.

\subsubsection{Summary}
\label{sec:summ}

In the preceding subsections, we have learned the following. Given corrections to $K$ of order $\delta \sim \alpha'^2/R^4$ and corrections to $\tW$ of order $\eta \sim e^{-\delta}$:
\begin{itemize}

\item Tree-level vacuum values of flux-modulus variables are corrected up to a scale \\${\cal C} = \delta|X|+\delta|Y|+\delta|Z|+\delta^{-1}\eta$.

\item Eigenvalues of the axion-dilaton/complex-structure mass matrix are corrected up to a scale ${\cal C'}= \delta(|X|+|Y|+|Z|)^2 + \delta^{-2}\eta(|X|+|Y|+|Z|)+\delta^{-3}\eta^2$.

\item Physically-reasonable, fully-supersymmetric vacua can (potentially) be built from supersymmetric tree-level vacua ($|Y|=0$) if $|X|\lesssim {\cal C}$. Two ways to achieve this are $|X|\lesssim \delta^{-1}\eta$ \eqref{KKLT} (\ie\ the KKLT scenario) and $|X|\lesssim \delta |Z|$ \eqref{KKLT2}.

- We can retain tree-level control over the variables $\Z{a}$.

- The cosmological constant is always AdS, and requires uplifting.

\item Consistent vacua that break supersymmetry with K\"ahler stabilization can be built from supersymmetric tree-level vacua ($|Y|=0$) in either of the two cases \eqref{KKLT} or \eqref{KKLT2}; or under the new condition $|X|\sim \delta^{-2}\eta$ \eqref{BBCQ} (\eg\ the large-volume scenario).

- This allows some tree-level control over $X$ as well as the $\Z{a}$.

- But unless $|X|\lesssim {\cal C}$, parametrically small (and positive) $\Lambda$ probably cannot be directly achieved.

\item Consistent non-supersymmetric vacua can be built from tree-level non-supersymmetric vacua ($|Y| \neq 0$) if $|Y|^2\lesssim \delta|X|^2$ \eqref{my1}.

- In principle, this can allow tree-level control over all the variables $X$, $\Y{i}$, and $\Z{a}$.

- A parametrically small cosmological constant (dS or AdS), controlled by the details of stringy corrections, is possible.

- But parametrically small $M_{susy}$ and $M_{3/2}$ are not possible unless $|X|\lesssim {\cal C}$ (\eg\ unless $|X|\lesssim \delta|Z|$), relinquishing control over the precise values of $X$ and the supersymmetry-breaking scale.

- Stability in axion-dilaton/complex-structure directions cannot be assured beyond tree level if any tree-level eigenvalues are $\lesssim |Y|^2$.

\end{itemize}

\subsection{Application to the models}
\label{sec:modelanal}

The above analysis can be applied very directly to the tree-level models of Section \ref{sec:MODELS}, since the varieties ${\cal X}_n$ in Section \ref{sec:X} (and Appendix \ref{app:n3sol}) tell us exactly which vacuum values the flux-modulus variables can take.

First consider the tree-level-supersymmetric vacua, \ie\ the `$\mb{S}$' branches, for any $n=1,2,3$. We claimed in Section \ref{sec:X} that they could always be extended to good solutions, and indeed it is always possible to satisfy conditions \eqref{KKLT}, \eqref{KKLT2}, or \eqref{BBCQ} because the $\Z{a}$ and $X$ are completely independent. Therefore, we can (potentially) use the models to form fully supersymmetric KKLT-like vacua, or the alternative vacua arising from \eqref{KKLT2}, with tree-level control over the $\Z{a}$; or nonsupersymmetric ``large volume'' vacua with more control over $X$ as well. The supersymmetry-breaking scale in the latter case will of course be controlled by the details of stringy corrections. As for stability, the eigenvalues of the $\mb{S}$ branches all depend on the $\Z{a}$, and can be made as large as desired by increasing the magnitudes of the $\Z{a}$. In particular, the eigenvalues can surpass the scale ${\cal C}'$, guaranteeing stability in the axion-dilaton and complex-structure directions.

Now consider the ``antisupersymmetric'' supersymmetry-breaking branches $\ol{\mb{S}}$. Since $|X|=|Z|=0$ along these branches for any $n$, it is almost impossible to satisfy condition \eqref{my1}, leading to an inconsistent cosmological constant. This was precisely the objection raised against these solutions in \cite{D&DI}. It \emph{is} possible to satisfy \eqref{my1} if $|Y|\lesssim {\cal C}$ (for example, if $|Y|\lesssim \delta^{-1}\eta$), but then we lose much predictive control over the tree-level solution. Thus, the antisupersymmetric solutions are perfectly good to avoid.

The situation is greatly improved, however, with ``intermediate'' nonsupersymmetric bran-ches. For any $n$, the common feature of these tree-level solutions is that the $\{\Y{i}\}$ and the $\{X,\Z{a}\}$ are both (mostly) nonzero, and are independently tunable. Therefore, it is always possible to satisfy condition \eqref{my1}, and to potentially extend to complete, consistent nonsupersymmetric vacua while retaining some tree-level control over all the flux-modulus variables. The resulting vacua could be dS or AdS, depending on the specific structure of corrections in \eqref{ns-Lambda}.

The scale of supersymmetry breaking and the gravitino mass of nonsupersymmetric vacua are given by \eqref{ns-susy} and \eqref{ns-grav}, respectively. As explained in Section \ref{sec:nonsusy-corr}, these parameters can be made parametrically small, which may be physically desirable, if $|X|\lesssim {\cal C}$. For the $n=1$ intermediate branches ($\mb{A}$ and $\mb{A'}$), $|X|\lesssim{\cal C}$ would force $|Z|\lesssim {\cal C}$ as well, because there is a single $\Z{i}$ and it is related to $X$ --- thus, for $n=1$ it is impossible to make $M_{susy}$ and $M_{3/2}$ parametrically small without completely losing control of the tree-level structure. For $n=2$ and $n=3$, however, there arise intermediate branches on which the $\Z{i}$ and $X$ are more independent: namely, branches $\mb{A}$, $\mb{A'}$, and $\mb{D}$ for $n=2$, and branches $\mb{A_i}$ and $\mb{C_i}$ for $n=3$ (in Appendix \ref{app:n3sol}). Therefore, for $n>1$, one could potentially construct physically-sensilble solutions which retain tree-level control over some of the $\Z{i}$. Solutions such as $\mb{A}$ and $\mb{A'}$ for $n=2$ and the $\mb{A_i}$ for $n=3$, which have the most freedom in the magnitudes of the $\Z{i}$, may be most useful in such a construction.

To evaluate stability for the intermediate branches, we can look at the explicit expressions for eigenvalues given in \eqref{n1evals}, \eqref{n2evals}, and \eqref{n3evals}. Unfortunately, every intermediate branch has an eigenvalue proportional to $|Y|^2$ (or some magnitudes of $\Y{i}$'s). Thus, by the argument in section \ref{sec:stab}, the intermediate branches can never be \emph{assured} stability beyond tree level because this eigenvalue will be $\lesssim {\cal C}'$. Nevertheless, it may still be possible to engineer stringy corrections such that complete stability is obtained.

\section{Some comments on potentials: $|DW|^2$ vs. $|DW|^2-3|W|^2$}
\label{sec:scale-noscale}

We finish with some comments about different ``choices'' of tree-level potentials which appear in the the literature. Several analyses of tree-level vacua, statistical and otherwise, have used the potential
\be V' = T_3\,(|D_{\hat{i}}\tW|^2-3|\tW|^2) \nno \ee
rather than the no-scale potential
\be V = T_3\,|D_{\hat{i}}\tW|^2 \nno \ee
to stabilize axion-dilaton and complex-structure moduli. This includes for example \cite{D&DI, D&DII}, as well as the more recent \cite{HMR, Soroush}. The initial motivation for using $V'$ in \cite{D&DI, D&DII} was to include a flavor of the dynamics of K\"ahler moduli and the possibility of an AdS cosmological constant without explicitly adding stringy corrections.%
\footnote{We thank F. Denef for communication on this subject.} %
We can attempt to reinterpret the use of $V'$ in light of our analysis of corrections from Section \ref{sec:analysis}.

First, observe that critical points of the two potentials $V,\,V'$ are in one-to-one correspondence. Indeed, if we compare the critical-point equations for $V$,
\bse
\begin{align}
 D_{\hat{0}} V &= Z_{\hat{c}} \bar{Y}^{\hat{c}} + Y_{\hat{0}}\bar{X} =0\,, \tag{\ref{DV2}a} \\
 D_{\hat{a}} V &= Z_{\hat{a}}\bar{Y}_{\hat{0}} + \cF_{\hat{a}\hat{b}\hat{c}}\bar{Z}^{\hat{b}}\bar{Y}^{\hat{c}}+Y_{\hat{a}}\bar{X} = 0\,, \tag{\ref{DV2}b}
\end{align}
\ese
to those for $V'$,
\bse
\begin{align}
\D0V' &= \Z{c}\bar{Y}^{\hat{c}}-2\Y0\cX\,, \\
\D{a}V' &= \Z{a}\cY0 +\Fab \bar{Z}^{\hat{b}}\bar{Y}^{\hat{c}}-2\Y{a}\cX\,,
\end{align}
\ese
in terms of flux-modulus variables, we see that they are simply related by a transformation
\be X \leftrightarrow -2X\,. \ee
Therefore, any abstract solution to $dV=0$ can be obtained from a solution to $dV'=0$ by setting $X\goesto-2X$, and vice versa. The main difference between $V$ and $V'$, however, is \emph{stability}: the extra $-3|W|^2$ tends to destabilize critical points of the potential $V'$.

Clearly, supersymmetric vacua of the no-scale potential are stable at tree level, and we have argued that their stability can be guaranteed beyond tree level as well. However, with potential $V'$, the mass matrix decomposition $M_{\mbc,\tau}^2=H^2$ from \eqref{M-decomp} gets replaced by \cite{D0,D&DI,Denef:2004dm}
\be M_{\mbc,\tau} = H^2-3|X|H\,, \ee
which leads to a new condition required%
\footnote{All supersymmetric vacua are AdS, and vacua in AdS space are protected from decay even without a (naively) positive-definite mass matrix \cite{BF}. However, tachyonic directions become relevant after a KKLT-like dS uplift.} %
for tree-level stability:
\be 2|X| \lesssim |Z_{\hat{a}}|\,. \label{newcond} \ee
For nonsupersymmetric vacua, the analysis of the mass matrix is much more complicated (see for example \cite{D&DII, Soroush}), but a version of \eqref{newcond} seems to remain true. Roughly, one must require that $X$ and the $\Z{a}$ have some degree of independence on a given branch of solutions in order to obtain supersymmetric vacua. Thus, for example, all $n=1$ supersymmetry-breaking vacua at large complex structure are unstable in $V'$, because they all have a constraint relating $|X|$ and $|\Z{1}|$. This was the reason that such vacua were precluded in \cite{D&DI}. Similarly, if we consider critical points of $V'$ corresponding to the $n=2$ supersymmetry-breaking branches in our models, we find that only the $\mb{A}$ and $\mb{A}'$ branches (which allow the most freedom between $|X|$ and the $|\Z{i}|$) contain stable vacua, and stability happens precisely when $2|X|<|Z|$.

The fact that $V'$ is unstable roughly when \eqref{newcond} is violated could be used to one's advantage. Considering the analysis of corrections in Section \ref{sec:approx}, we see that condition \eqref{newcond} is somewhat similar to our mechanisms for ``swamping out'' $|X|$ as in \eqref{KKLT2} and \eqref{XZswamp}. For supersymmetric vacua, \eqref{KKLT2} was useful in allowing a small cosmological constant, and could also ensure stability of a vacuum beyond tree level, since eigenvalues of $M_{\mbc,\tau}^2$ tend to grow with increasing $|Z|^2$. For nonsupersymmetric vacua, $|X|\ll|Z|$ allowed a low scale of supersymmetry breaking. Therefore, when performing a \emph{statistical} analysis, using the tree-level potential $V'$ and excluding ``unstable'' vacua may actually be a good way to approximately restrict the configuration space to models with desirable physical properties. One should keep in mind, however, that actual stability information coming from $V'$ is not physical. In order to analyze a specific stabilization model, using the no-scale potential together with some consideration of stringy corrections is still the appropriate approach.

\section{Conclusions}

We have presented several explicit, computable models of tree-level moduli stabilization near large complex structure in type IIB orientifold compactifications. The unifying and simplifying feature of our models was the presence of a single nonvanishing Yukawa coupling near large complex structure. This restricted our possible models to $n=1,$ 2, and 3 complex-structure moduli. Using the formalism of flux-modulus variables of \cite{D&DI}, we were able to give explicit, abstract descriptions of the solutions to $dV=0$, both supersymmetric and nonsupersymmetric. We also showed that given a specific compactification manifold there is enough freedom in our models to create a desired abstract vacuum at any point in its large-complex-structure moduli space, up to subtleties of flux quantization.

At tree level, all the vacua of the no-scale potential are automatically stable for our one-Yukawa, large-complex-structure models. That is, the axion-dilaton/complex-structure mass matrix is positive-definite. However, to properly evaluate the stability and other physical properties of vacua, it is necessary to go beyond tree level. As such, we performed a simple but general analysis of how stringy corrections can effect tree-level structure. We reclassified some popular stabilization scenarios such as KKLT and the large-volume vacua of \cite{BBCQ}, and also found some new possibilities for constructing complete, controllable stabilizations from tree-level vacua. In particular, we found that it is (in principle) possible to build consistent vacua from nonsupersymmetric tree-level solutions, provided that the scale of supersymmetric breaking is not too high. We can realize these scenarios in the axion-dilaton/complex-structure sector using the intermediate supersymmetry-breaking branches of our $n>1$ models. These nonsupersymmetric vacua are potentially interesting because they can have a positive cosmological constant without resorting to extra uplifting mechanisms. Unfortunately, if a parametrically low supersymmetry-breaking scale is required in the hidden sector, we can still build consistent models from nonsupersymmetric tree-level solutions, but we necessarily lose some control over the precise scale of supersymmetry breaking and the magnitude of the tree-level (vacuum) superpotential. The resulting vacua also tend to have modulino masses significantly greater than the gravitino mass.

Our analysis of models and stringy corrections is based on using the no-scale form of the scalar potential at tree level, which is most appropriate for an honest compactification. However, in light of our study of corrections, we attempted to provide additional motivation for the use of another form of the scalar potential in \emph{statistical} analyses, as is done for example in \cite{D&DI, D&DII, HMR}. Ignoring the no-scale cancellation causes some vacua with undesirable physical properties to become destabilized in the alternative ``$-3|W|^2$'' potential, and could be an effective way to restrict the configuration space.

We hope that our explicit constructions may be useful in other studies of tree-level vacua, such as investigations of paths and instantons in the flux landscape \cite{Chialva:2007sv, Johnson:2008kc}. It would also be interesting to use the explicit form of stringy corrections to the K\"ahler potential and superpotential to actually realize some of the unexplored stabilization scenarios of Section \ref{sec:analysis} --- though some of these scenarios may require concurrent stabilization of both K\"ahler and complex-structure moduli, making them difficult to analyze. Additionally, it could be fruitful to extend the models in this paper to the open-string sector, in particular generalizing the use of flux-modulus variables to superpotentials derived from F-theory, which include D7 moduli. We hope to address some of these issues in future work.

\newpage
\noindent\textbf{\large Acknowledgements}
\vspace{.1cm}

It is a pleasure to thank H. Ooguri for inspiring and advising this project, as well as J. Conlon, F. Denef, M. Dine, S. Gukov, M. Johnson, J. Kumar, M. Larfors, J. Marsano, C. Melby-Thompson, F. Quevedo, K. Saraikin, N. Saulina, and S. Schafer-Nameki for many insightful discussions. The author would also like to thank IPMU, Tokyo, where part of this work was completed, for their great hospitality. This work was supported in part by DOE grant DE-FG02-92ER40701, the World Premier International Research Center Initiative of MEXT of Japan, and a National Defense Science and Engineering Graduate Fellowship.

\vspace{.5cm}

\appendix

\section{Details of maps to (and from) flux-modulus variables}
\label{app:mapdata}

This appendix complements Section \ref{sec:maps}, providing more details of the constructions decribed there.

Recall that we defined the integrals of the complexified flux as
\begin{subequations} \label{app-flux}
\begin{align}
M^i &= \int_{A_i}G_3 = (m_{RR}^i-\tau m_{NSNS}^i)\,, \\
E_i &= \int_{B^i}G_3 = (e^{RR}_i-\tau e^{NSNS}_i)\,,
\end{align}
\end{subequations}
where
\begin{subequations}
\begin{align} 
&m_{RR}^i = \int_{A_i}F_{RR}\,, \qquad e^{RR}_i = \int_{B^i}F_{RR}\,, \\
&m_{NSNS}^i = \int_{A_i}H_{NSNS}\,, \qquad e^{NSNS}_i = \int_{B^i}H_{NSNS}\,.
\end{align}
\end{subequations}
Note that the real flux integrals can be easily obtained from the complex flux integrals at any finite string coupling $\Im\,\tau>0$; for example
\be m^i_{NSNS} = -\frac{\Im\,M^i}{\Im\,\tau}\,,\qquad m^i_{RR} = \Re\,M^i-\Re\,\tau\frac{\Im\,M^i}{\Im\,\tau}\,. \label{mM} \ee

The superpotential, as in \eqref{large_W}, is
\begin{align} W
&= -\frac{1}{6}M^0y_{abc}t^at^bt^c+\frac{1}{2}y_{abc}M^at^bt^c+(q_{ab}M^a+\ell_bM^0+E_b)t^b+(\ell_aM^a+2cM^0+E_0)\,. \label{large_W-ap}
\end{align}

We will need some information about the axion-dilaton moduli space. The K\"ahler potential $K_\tau = -\log[-i(\tau-\bar{\tau})]$ leads to
\begin{subequations} \label{taugeom}
\be \pd_0K = -\frac{1}{\tau-\bar{\tau}}\,, \qquad g_{0\bar{0}}=-\frac{1}{(\tau-\bar{\tau})^2}\,, \qquad \Gamma^0_{00} = -\frac{2}{\tau-\bar{\tau}} \ee
\be e{_{\hat{0}}}^0 = -(\tau-\bar{\tau})\,, \ee
\end{subequations}
and
\be D_{\hat{0}}W = [1-(\tau-\bar{\tau})\pd_0]W\,.  \label{D0} \ee
Since the complex structure of the coefficients in $W$ is induced from \eqref{app-flux}, we find that $\D{0}$ always acts on these coefficients by complex conjugation.

\subsection{n=1}

In the case $n=1$, the nonvanishing Yukawa coupling is $y_{111}$. The superpotential \eqref{large_W-ap} may be written as
\be  W = At^3 + B t^2 + Ct + D\,,  \label{n1W} \ee
with
\be \left(\begin{array}{c} A \\ B \\ C \\ D \end{array}\right) =
 \left(\begin{array}{cccc}
  -\frac{1}{6}y_{111} & 0 & 0 & 0 \\
  0 & \frac{1}{2}y_{111} & 0 & 0 \\
  \ell_1 & q_{11} & 0 & 1 \\
  2c & \ell_1 & 1 & 0 \end{array}\right)
  \left(\begin{array}{c} M^0 \\ M^1 \\ E_0 \\ E_1 \end{array}\right).
\ee
This is a nonsingular transformation when $y_{111}\neq 0$, and can be inverted as
\be \left(\begin{array}{c} M^0 \\ M^1 \\ E_0 \\ E_1 \end{array}\right)
 = \frac{2}{y_{111}}\left(\begin{array}{cccc}
  -3 & 0 & 0 & 0 \\
  0 & 1 & 0 & 0 \\
  6 c & -\ell_1 & 0 & \frac{1}{2}y_{111} \\
  3 \ell_1 & -q_{11} & \frac{1}{2}y_{111} & 0 \end{array}\right)
 \left(\begin{array}{c} A \\ B \\ C \\ D \end{array}\right). \label{n1MA}
\ee

From \eqref{n1geom}, we see that the complex-structure covariant derivative acts on $W$ as 
\be D_{\hat{1}}W = \frac{1}{\sqrt{3}}[3-(t-\bar{t})\pd_t]W\,. \ee
The flux-modulus variables can then be computed as
\begin{align*}
 e^{-K/2}X=W &= At^3+Bt^2+Ct+D\,, \\
 e^{-K/2}Y_{\hat{0}}=D_{\hat{0}}W &= \bar{A}t^3+\bar{B}t^2+\bar{C}t+\bar{D}\,, \\
 e^{-K/2}Y_{\hat{1}}=D_{\hat{1}}W &= \frac{1}{\sqrt{3}}\left[3At^2\bar{t}+B(t^2+2|t|^2)+C(2t+\bar{t})+3D\right], \\
 e^{-K/2}Z_{\hat{1}}=D_{\hat{0}}D_{\hat{1}}W &= \frac{1}{\sqrt{3}}\left[3\bar{A}t^2\bar{t}+\bar{B}(t^2+2|t|^2)+\bar{C}(2t+\bar{t})+3\bar{D}\right],
\end{align*}
or, equivalently,
\be \left(\begin{array}{c} X \\ \sqrt{3}Y_{\hat{1}} \\ \bar{Y}_{\hat{\bar{0}}} \\ \sqrt{3}\bar{Z}_{\hat{\bar{1}}} \end{array}\right) =
 e^{K/2}\left(\begin{array}{cccc}
 t^3 \,&\, t^2 \,&\, t \,&\, 1 \\
 3t^2\bar{t} \,&\, t^2+2|t|^2 \,&\, 2t+\bar{t} \,&\, 3 \\
 \bar{t}^3 \,&\, \bar{t}^2 \,&\, \bar{t} \,&\, 1 \\
 3\bar{t}^2t \,&\, \bar{t}^2+2|t|^2 \,&\, 2\bar{t}+t \,&\, 3 \end{array}\right)
\left(\begin{array}{c} A \\ B \\ C \\ D \end{array}\right). \label{n1XA-app}
\ee
The matrix here is nonsingular as long as $\Im\,t> 0$, which should certainly hold at large complex structure, so we can invert \eqref{n1XA-app} as
\be \left(\begin{array}{c} A \\ B \\ C \\ D \end{array}\right) 
 = \frac{e^{-K/2}}{(t-\bar{t})^3}\,{\cal T}_1 \left(\begin{array}{c}
  X \\ \sqrt{3}Y_{\hat{1}} \\ \bar{Y}_{\hat{\bar{0}}} \\ \sqrt{3}\bar{Z}_{\hat{\bar{1}}}
  \end{array}\right), \label{n1AX}
\ee
with
\be {\cal T}_1 = \left(
\begin{array}{cccc}
 1 \,&\, -1 \,&\, -1 \,&\, 1 \\
 -3\bar{t} \,&\, 2\bar{t}+t \,&\, 3t \,&\, -\bar{t}-2t \\
 3\bar{t}^2 \,&\, -\bar{t}^2-2|t|^2 \,&\, -3t^2 \,&\, t^2+2|t|^2 \\
 -3\bar{t}^3 \,&\, t|t|^2 \,&\, t^3 \,&\, -\bar{t}|t|^2
\end{array}
\right). \label{n1AX2}
\ee

By combining the inverse expressions \eqref{n1AX}, \eqref{n1MA}, and \eqref{mM}, we see how any desired value of $(X,Y_{\hat{0}},Y_{\hat{1}},Z_{\hat{1}})$ may in principle be attained at fixed $(\tau,t)$ by tuning the eight real fluxes $(m^i_{RR},m^i_{NSNS},e^{RR}_i,e^{NSNS}_i)$. This construction agrees fully with the more direct computation suggested below \eqref{gen_map}. Note that the prefactor in \eqref{n1AX} can be written as $e^{-K/2}(t-\bar{t})^{-3}=i\frac{y_{111}}{6}e^{-K/2+K_\mbc}$. 

\subsection{n=2}

For $n=2$, with $y_{112}\neq 0$, the superpotential \eqref{large_W-ap} can be written
\be  W = A t_1^2 t_2 + B t_1^2 + C t_1t_2 + D t_1 + E t_2 + G\,, \label{n2W} \ee
with
\be \left( \begin{array}{c} A \\ B \\ C \\ D \\ E \\ G \end{array}\right)  = \left( \begin{array}{cccccc}
 -\frac{1}{2}y_{112} &0&0&0&0&0 \\
 0&0& \frac{1}{2}y_{112} &0&0&0 \\
 0& y_{112} &0&0&0&0 \\
 \ell_1 & q_{11} & q_{12} & 0 & 1 & 0 \\
 \ell_2 & q_{21} & q_{22} & 0 & 0 & 1 \\
 2c & \ell_1 & \ell_2 & 1 & 0 & 0 \end{array} \right)
\left( \begin{array}{c} M^0 \\ M^1 \\ M^2 \\ E_0 \\ E_1 \\ E_2 \end{array} \right), \label{n2AM}
\ee
or (since $y_{112}\neq0$)
\be \left( \begin{array}{c} M^0 \\ M^1 \\ M^2 \\ E_0 \\ E_1 \\ E_2 \end{array} \right) = {y_{112}}^{-1}\left(\begin{array}{cccccc}
 -2&0&0&0&0&0 \\ 0&0&1&0&0&0 \\ 0&2&0&0&0&0 \\
 4c & -2\ell_2 & -\ell_1 & 0&0&y_{112} \\
 2\ell_1 & -2q_{12} & -q_{11} & y_{112} & 0 & 0 \\
 2\ell_2 & -2q_{22} & -q_{12} & 0 & y_{112} & 0 \end{array}\right)
\left(\begin{array}{c} A \\ B \\ C \\ D \\ E \\ G \end{array} \right). \label{n2MA}
\ee

The complex-structure covariant derivatives are
\be D_{\hat{1}}W = \frac{1}{\sqrt{2}}[2-(t_1-\bar{t}_1)\pd_1]W \,,\qquad D_{\hat{2}}W = [1-(t_2-\bar{t}_2)\pd_2]W\,, \ee
whence $X=e^{K/2}W$, $Y_{\hat{i}}=e^{K/2}D_{\hat{i}}W$, and $Z_{\hat{a}}=e^{K/2}D_{\hat{0}}D_{\hat{a}}W$ are expressed as
\be \left( \begin{array}{c} X \\ \sqrt{2}\Y1 \\ \Y2 \\ \cY0 \\ \sqrt{2}\cZ1 \\ \cZ2 \end{array} \right) = e^{K/2}
\left(\begin{array}{cccccc}
 t_1^2 t_2 \,&\, t_1^2 \,&\, t_1 t_2 \,&\, t_1 \,&\, t_2 \,&\, 1 \\
  2|t_1|^2t_2 \,&\, 2|t_1|^2 \,&\, 2(\Re\,t_1) t_2 \,&\, 2\,\Re\,t_1 \,&\, 2t_2 \,&\, 2 \\
 t_1^2 \ol{t}_2 \,&\, t_1^2 \,&\, t_1 \ol{t}_2 \,&\, t_1 \,&\, \ol{t}_2 \,&\, 1 \\
 \ol{t}_1^2 \ol{t}_2 \,&\, \ol{t}_1^2 \,&\, \ol{t}_1 \ol{t}_2 \,&\, \ol{t}_1 \,&\, \ol{t}_2 \,&\, 1 \\
  2|t_1|^2\ol{t}_2 \,&\, 2|t_1|^2 \,&\, 2(\Re\,t_1) \ol{t}_2 \,&\, 2\,\Re\,t_1 \,&\, 2\ol{t}_2 \,&\, 2 \\
 \ol{t}_1^2 t_2 \,&\, \ol{t}_1^2 \,&\, \ol{t}_1 t_2 \,&\, \ol{t}_1 \,&\, t_2 \,&\, 1
\end{array} \right)
\left( \begin{array}{c} A \\ B \\ C \\ D \\ E \\ G \end{array} \right).
\label{n2XA}
\ee
Nonsingularity of this matrix only requires $\Im\,t_1,\,\Im\,t_2>0$ (satisfied at large complex structure), and the inverse relation is
\begin{align} \left( \begin{array}{c} A \\ B \\ C \\ D \\ E \\ G \end{array} \right) &= 
i\frac{y_{112}}{2}e^{-K/2+K_\mbc}\,
 \,{\cal T}_2
 \left( \begin{array}{c} X \\ \sqrt{2}\Y1 \\ \Y2 \\ \cY0 \\ \sqrt{2}\cZ1 \\ \cZ2 \end{array} \right), \label{n2AX}
\end{align}
\be {\cal T}_2 = \left( \begin{array}{cccccc}
  1 & -1 & -1 & -1 & 1 & 1 \\
  -\ol{t}_2 & \ol{t}_2 & t_2 & t_2 & -t_2 & -\ol{t}_2 \\
  -2\ol{t}_1 & 2\,\Re\,t_1 & 2\ol{t}_1 & 2t_1 & -2\,\Re\,t_1 & -2t_1 \\
  2\ol{t}_1 \ol{t}_2 & -2(\Re\,t_1)\ol{t}_2 & -2\ol{t}_1 t_2 & -2 t_1 t_2 & 2(\Re\,t_1)t_2 & 2t_1 \ol{t}_2 \\
  \ol{t}_1^2 & -|t_1|^2 & -\ol{t}_1^2 & -t_1^2 & |t_1|^2 & t_1^2 \\
  -\ol{t}_1^2\ol{t}_2 & |t_1|^2\ol{t}_2 & \ol{t}_1^2 t_2 & t_1^2 t_2 & -|t_1|^2 t_2 & -\ol{t}_2t_1^2
  \end{array} \right). \label{n2AX2}
\ee
Combining \eqref{n2AX}, \eqref{n2MA}, and \eqref{mM} gives the prescription for achieving desired $(X,Y_{\hat{i}},Z_{\hat{a}})$ at fixed $(\tau,t_1,t_2)$ by tuning the fluxes.

\subsection{n=3}

For $n=3$ and $y_{123}\neq 0$ the potential is
\be W = At_1t_2t_3 + Bt_1t_2+Ct_2t_3+Dt_1t_3 + Et_1 + Gt_2 + Ht_3 + I\,, \ee
with
\be \left(\begin{array}{c} A \\ B \\ C \\ D \\ E \\ G \\ H \\ I \end{array}\right) = 
 \left(\begin{array}{cccccccc}
  -y_{123} & 0 & 0 & 0 & 0 & 0 & 0 & 0 \\
  0 & 0 & 0 & y_{123} & 0 & 0 & 0 & 0 \\
  0 & y_{123} & 0 & 0 & 0 & 0 & 0 & 0 \\
  0 & 0 & y_{123} & 0 & 0 & 0 & 0 & 0 \\
  \ell_1 & q_{11} & q_{12} & q_{13} & 0 & 1 & 0 & 0 \\
  \ell_2 & q_{21} & q_{22} & q_{23} & 0 & 0 & 1 & 0 \\
  \ell_3 & q_{31} & q_{32} & q_{33} & 0 & 0 & 0 & 1 \\
  2c & \ell_1 & \ell_2 & \ell_3 & 1 & 0 & 0 & 0
 \end{array}\right)
 \left(\begin{array}{c} M^0 \\ M^1 \\ M^2 \\ M^3 \\ E_0 \\ E_1 \\ E_2 \\ E_3 \end{array}\right), \label{n3AM}
\ee
or
\be \left(\begin{array}{c} M^0 \\ M^1 \\ M^2 \\ M^3 \\ E_0 \\ E_1 \\ E_2 \\ E_3 \end{array}\right) =
 \frac{1}{y_{123}}\left(\begin{array}{cccccccc}
 -1 & 0 & 0 & 0 & 0 & 0 & 0 & 0 \\
 0 & 0 & 1 & 0 & 0 & 0 & 0 & 0 \\
 0 & 0 & 0 & 1 & 0 & 0 & 0 & 0 \\
 0 & 1 & 0 & 0 & 0 & 0 & 0 & 0 \\
 2c & -\ell_3 & -\ell_1 & -\ell_2 & 0 & 0 & 0 & y_{123} \\
 \ell_1 & -q_{13} & -q_{11} & -q_{12} & y_{123} & 0 & 0 & 0 \\
 \ell_2 & -q_{23} & -q_{21} & -q_{22} & 0 & y_{123} & 0 & 0 \\
 \ell_3 & -q_{33} & -q_{31} & -q_{32} & 0 & 0 & y_{123} & 0
 \end{array}\right)
 \left(\begin{array}{c} A \\ B \\ C \\ D \\ E \\ G \\ H \\ I \end{array}\right). \label{n3MA}
\ee

The covariant derivatives for $a=1,2,3$ are all
\be D_{\hat{a}}W = [1-(t_a-\bar{t}_a)\pd_a]W \qquad\mbox{(no sum over $a$)}\,, \label{n3D} \ee
giving
\be \left(\begin{array}{c} X \\ Y_{\hat{1}} \\ Y_{\hat{2}} \\ Y_{\hat{3}} \\ \bar{Y}_{\hat{\bar{0}}} \\ \bar{Z}_{\hat{\bar{1}}} \\ \bar{Z}_{\hat{\bar{2}}} \\ \bar{Z}_{\hat{\bar{3}}} \end{array}\right)
 = e^{K/2}
 \left(
\begin{array}{llllllll}
 t_1 {t_2} {t_3} \,&\, t_1 {t_2} \,&\, {t_2} {t_3} \,&\,
   t_1 {t_3} \,&\, t_1 \,&\, {t_2} \,&\, {t_3} \,&\, 1 \\
 \bar{t}_1 {t_2} {t_3} \,&\, \bar{t}_1 {t_2} \,&\, {t_2} {t_3} \,&\,
   \bar{t}_1 {t_3} \,&\, \bar{t}_1 \,&\, {t_2} \,&\, {t_3} \,&\, 1 \\
 {\bar{t}_2} t_1 {t_3} \,&\, {\bar{t}_2} t_1 \,&\, {\bar{t}_2} {t_3} \,&\,
   t_1 {t_3} \,&\, t_1 \,&\, {\bar{t}_2} \,&\, {t_3} \,&\, 1 \\
 {\bar{t}_3} t_1 {t_2} \,&\, t_1 {t_2} \,&\, {\bar{t}_3} {t_2} \,&\,
   {\bar{t}_3} t_1 \,&\, t_1 \,&\, {t_2} \,&\, {\bar{t}_3} \,&\, 1 \\
 \bar{t}_1 {\bar{t}_2} {\bar{t}_3} \,&\, \bar{t}_1 {\bar{t}_2} \,&\, {\bar{t}_2} {\bar{t}_3} \,&\,
   \bar{t}_1 {\bar{t}_3} \,&\, \bar{t}_1 \,&\, {\bar{t}_2} \,&\, {\bar{t}_3} \,&\, 1\\
  t_1 {\bar{t}_2} {\bar{t}_3} \,&\,  t_1{\bar{t}_2} \,&\, {\bar{t}_2} {\bar{t}_3} \,&\,
    t_1{\bar{t}_3} \,&\, t_1 \,&\, {\bar{t}_2} \,&\, {\bar{t}_3} \,&\, 1 \\
 \bar{t}_1 {t_2} {\bar{t}_3}  \,&\, \bar{t}_1 {t_2} \,&\, {t_2}{\bar{t}_3}  \,&\,
   \bar{t}_1 {\bar{t}_3} \,&\, \bar{t}_1 \,&\, {t_2} \,&\, {\bar{t}_3} \,&\, 1 \\
 \bar{t}_1 {\bar{t}_2} {t_3} \,&\, \bar{t}_1 {\bar{t}_2} \,&\, {\bar{t}_2} {t_3} \,&\,
   \bar{t}_1 {t_3} \,&\, \bar{t}_1 \,&\, {\bar{t}_2} \,&\, {t_3} \,&\, 1
\end{array}\right)
\left(\begin{array}{c} A \\ B \\ C \\ D \\ E \\ G \\ H \\ I \end{array}\right). \label{n3XA}
\ee
The inverse relation is
\be \left(\begin{array}{c} A \\ B \\ C \\ D \\ E \\ G \\ H \\ I \end{array}\right) 
 = 
 i\,y_{123}\,e^{-K/2+K_\mbc}
 \,{\cal T}_3
 \left(\begin{array}{c} X \\ Y_{\hat{1}} \\ Y_{\hat{2}} \\ Y_{\hat{3}} \\ \bar{Y}_{\hat{\bar{0}}} \\ \bar{Z}_{\hat{\bar{1}}} \\ \bar{Z}_{\hat{\bar{2}}} \\ \bar{Z}_{\hat{\bar{3}}} \end{array}\right), \label{n3AX}
\ee
\be {\cal T}_3 =
 \left(
\begin{array}{llllllll}
 -1 & 1 & 1 & 1 & 1 & -1 & -1 & -1 \\
 {\bar{t}_3} & -{\bar{t}_3} & -{\bar{t}_3} & -{t_3} & -{t_3} & {t_3} &
   {t_3} & {\bar{t}_3} \\
 {\bar{t}_1} & -{t_1} & -{\bar{t}_1} & -{\bar{t}_1} & -{t_1} & {\bar{t}_1} &
   {t_1} & {t_1} \\
 {\bar{t}_2} & -{\bar{t}_2} & -{t_2} & -{\bar{t}_2} & -{t_2} & {t_2} &
   {\bar{t}_2} & {t_2} \\
 -{\bar{t}_2} {\bar{t}_3} & {\bar{t}_2} {\bar{t}_3} & {\bar{t}_3} {t_2} & {\bar{t}_2}
   {t_3} & {t_2} {t_3} & -{t_2} {t_3} & -{\bar{t}_2} {t_3} &
   -{\bar{t}_3} {t_2} \\
 -{\bar{t}_1} {\bar{t}_3} & {\bar{t}_3} {t_1} & {\bar{t}_1} {\bar{t}_3} & {\bar{t}_1}
   {t_3} & {t_1} {t_3} & -{\bar{t}_1} {t_3} & -{t_1} {t_3} &
   -{\bar{t}_3} {t_1} \\
 -{\bar{t}_1} {\bar{t}_2} & {\bar{t}_2} {t_1} & {\bar{t}_1} {t_2} & {\bar{t}_1}
   {\bar{t}_2} & {t_1} {t_2} & -{\bar{t}_1} {t_2} & -{\bar{t}_2} {t_1}
   & -{t_1} {t_2} \\
 {\bar{t}_1} {\bar{t}_2} {\bar{t}_3} & -{\bar{t}_2} {\bar{t}_3} {t_1} & -{\bar{t}_1}
   {\bar{t}_3} {t_2} & -{\bar{t}_1} {\bar{t}_2} {t_3} & -{t_1} {t_2}
   {t_3} & {\bar{t}_1} {t_2} {t_3} & {\bar{t}_2} {t_1} {t_3} &
   {\bar{t}_3} {t_1} {t_2}
\end{array}\right). \label{n3AX2}
\ee

Again, combining $\eqref{n3AX}$ and \eqref{n3MA} with \eqref{mM} produces the desired prescription for tuning fluxes.

\section{Classifying solutions for $n=3$}
\label{app:n3sol}

Finally, we look at the abstract solutions to $dV=0$ for out models in the case $n=3$; \ie\ we examine the abstract variety ${\cal X}_3$. This appendix complements Section \ref{sec:X}.

Since the nonvanishing Yukawa coupling $\cF_{\hat{1}\hat{2}\hat{3}}=1$ preserves a permutation symmetry of the indices $(\hat{1},\hat{2},\hat{3})$, the critical-point equations and the set of solutions in this case also preserve this symmetry.
The critical-point equations are
\bse
\begin{align}
 D_{\hat{0}}V &= \Z1\cY1+\Z2\cY2+\Z3\cY3+\cX\Y0 = 0\,, \\
 D_{\hat{1}}V &= \Z1\cY0+\cY3\cZ2+\cY2\cZ3+\cX\Y1 = 0\,, \\
 D_{\hat{2}}V &= \Z2\cY0+\cY3\cZ1+\cY1\cZ3+\cX\Y2 = 0\,, \\
 D_{\hat{3}}V &= \Z3\cY0+\cY2\cZ1+\cY1\cZ2+\cX\Y3 = 0\,.
\end{align} \label{n3eqs}
\end{subequations}
As for $n=1$ and $n=2$, the solutions can again be parametrized by a collection of magnitudes and phases. To conserve space, we can give the dependence on phases separately, since it is the same for every solution; in terms of free parameters $(\alpha,\beta,\gamma,\delta)$, we have
\begin{align}
\arg\,X = 2\alpha\,, &\qquad \arg\,\Y0 = \alpha+\beta+\gamma+\delta\,, \\
\arg\,\Z1 = 2\beta\,, &\qquad \arg\,\Y1 = \alpha+\beta-\gamma-\delta\,, \\
\arg\,\Z2 = 2\gamma\,, &\qquad \arg\,\Y2 = \alpha-\beta+\gamma-\delta\,, \\
\arg\,\Z3 = 2\delta\,, &\qquad \arg\,\Y3 = \alpha-\beta-\gamma+\delta\,.
\end{align}
The various branches of solutions are then described by relations among the magnitudes of the flux-modulus variables, as shown in Table \ref{tab:n3sol}.
%
%
\begin{table}[hbt]
\centering
\small
$\begin{array}{ccccccccc}
\mbox{Branch} & |X| & |\Z1| & |\Z2| & |\Z3| & |\Y0| & |\Y1| & |\Y2| & |\Y3| 
   \vspace{.05cm}\\ \hline  \vspace{-.4cm}\\
\mb{S} & \xi & \zeta_1 & \zeta_2 & \zeta_3 & 0 & 0 & 0 & 0 \vspace{.15cm} \\
\mb{\bar{S}} &  0 & 0 & 0 & 0 & \upsilon_0 & \upsilon_1 & \upsilon_2 & \upsilon_3 \vspace{.15cm} \\
\mb{A_1},\mb{A_2} &  \xi  & \zeta_1  & \zeta_2  & -(\xi\pm\zeta_1\pm\zeta_2) & \upsilon & \pm\upsilon  & \pm\upsilon  & \upsilon 
 \vspace{.15cm} \\
\mb{B_1},\mb{B_2} & \xi &\pm\xi & \pm\xi & \xi & \upsilon_0 & \upsilon_1 & \upsilon_2 & -(\upsilon_0\pm\upsilon_1\pm\upsilon_2)  \vspace{.15cm} \\
\mb{C_1},\mb{C_2} & \xi & \zeta & \pm\xi & \pm\zeta & \upsilon_0 & \upsilon_1 & \mp\upsilon_0 & \mp\upsilon_1
\end{array} $
\caption{Solutions to $dV=0$ for $n=3$, up to permutations of $(\hat{1},\hat{2},\hat{3})$}
\label{tab:n3sol}
\end{table}

\noindent %
Permutations of $(\hat{1},\hat{2},\hat{3})$ will produce two more $\mb{A}$-branches, two more $\mb{B}$-branches, and two more $\mb{C}$-branches. Negative magnitudes are to be understood as changing the phase of a variable by $\pi$; \emph{cf}. Table \ref{tab:n2sol}.

The eigenvalues of the orthonormal-frame mass matrix corresponding to each of these solutions are
\bse \label{n3evals}
\begin{align}
\mb{S}: &\quad (\xi\pm\zeta_1\pm\zeta_2\pm\zeta_3)^2,\,\,(\xi\mp\zeta_1\pm\zeta_2\pm\zeta_3)^2,\,\,\nno\\ &\qquad (\xi\pm\zeta_1\mp\zeta_2\pm\zeta_3)^2,\,\,(\xi\pm\zeta_1\pm\zeta_2\mp\zeta_3)^2 \vspace{.2cm} \\
\mb{S}: &\quad (\upsilon_0\pm\upsilon_1\pm\upsilon_2\pm\upsilon_3)^2,\,\,(\upsilon_0\mp\upsilon_1\pm\upsilon_2\pm\upsilon_3)^2,\,\,\nno\\ &\qquad (\upsilon_0\pm\upsilon_1\mp\upsilon_2\pm\upsilon_3)^2,\,\,(\upsilon_0\pm\upsilon_1\pm\upsilon_2\mp\upsilon_3)^2 \vspace{.2cm} \\
%
\mb{A_1}: &\quad 4(\xi+\zeta_1+\zeta_2)^2+4\upsilon^2,\,\, 4(\xi+\zeta_1)^2,\,\, 4(\xi+\zeta_2)^2,\,\,4(\zeta_1+\zeta_2)^2,\,\, \nno\\&\qquad  4(\xi^2+\upsilon^2),\,\,4(\zeta_1^2+\upsilon^2),\,\,4(\zeta_2^2+\upsilon^2),\,\,16\upsilon^2
\vspace{.25cm} \\
\mb{A_2}: &\quad 4(\xi-\zeta_1-\zeta_2)^2+4\upsilon^2,\,\, 4(\xi-\zeta_1)^2,\,\, 4(\xi-\zeta_2)^2,\,\,4(\zeta_1+\zeta_2)^2,\,\, \nno\\&\qquad  4(\xi^2+\upsilon^2),\,\,4(\zeta_1^2+\upsilon^2),\,\,4(\zeta_2^2+\upsilon^2),\,\,16\upsilon^2
\vspace{.25cm} \\
\mb{B_1}: &\quad 4(\upsilon_0+\upsilon_1+\upsilon_2)^2+4\xi^2,\,\, 4(\upsilon_0+\upsilon_1)^2,\,\, 4(\upsilon_0+\upsilon_2)^2,\,\,4(\upsilon_1+\upsilon_2)^2,\,\, \nno\\&\qquad  4(\upsilon_0^2+\xi^2),\,\,4(\upsilon_1^2+\xi^2),\,\,4(\upsilon_2^2+\xi^2),\,\,16\xi^2
\vspace{.25cm} \\
\mb{B_2}: &\quad 4(\upsilon_0-\upsilon_1-\upsilon_2)^2+4\xi^2,\,\, 4(\upsilon_0-\upsilon_1)^2,\,\, 4(\upsilon_0-\upsilon_2)^2,\,\,4(\upsilon_1+\upsilon_2)^2,\,\, \nno\\&\qquad  4(\upsilon_0^2+\xi^2),\,\,4(\upsilon_1^2+\xi^2),\,\,4(\upsilon_2^2+\xi^2),\,\,16\xi^2
\vspace{.25cm} \\
\mb{C_1},\mb{C_2}: &\quad 4(\xi\pm\zeta)^2,\,\, 4(\upsilon_0\pm\upsilon_1)^2,\,\, 4(\xi^2+\upsilon_0^2),\,\, 4(\xi^2+\upsilon_1^2),\,\,4(\zeta^2+\upsilon_0^2),\,\, 4(\zeta^2+\upsilon_1^2)\,.
\end{align}
\ese

\vspace{.5cm}

\bibliographystyle{JHEP}
\bibliography{flux}

\providecommand{\href}[2]{#2}\begingroup\raggedright\begin{thebibliography}{10}

\bibitem{GKP}
S.~B. Giddings, S.~Kachru, and J.~Polchinski, {\it {Hierarchies from Fluxes in
  String Compactifications}},  {\em Phys. Rev.} {\bf D66} (2002) 106006,
  [\href{http://xxx.lanl.gov/abs/hep-th/0105097}{{\tt hep-th/0105097}}].

\bibitem{Denef}
F.~Denef, {\it {Les Houches Lectures on Constructing String Vacua}},
  \href{http://xxx.lanl.gov/abs/0803.1194}{{\tt arXiv:0803.1194}}.

\bibitem{KKLT}
S.~Kachru, R.~Kallosh, A.~Linde, and S.~P. Trivedi, {\it {De Sitter Vacua in
  String Theory}},  {\em Phys. Rev.} {\bf D68} (2003) 046005,
  [\href{http://xxx.lanl.gov/abs/hep-th/0301240}{{\tt hep-th/0301240}}].

\bibitem{Denef:2004dm}
F.~Denef, M.~R. Douglas, and B.~Florea, {\it {Building a Better Racetrack}},
  {\em JHEP} {\bf 06} (2004) 034,
  [\href{http://xxx.lanl.gov/abs/hep-th/0404257}{{\tt hep-th/0404257}}].

\bibitem{Denef:2005mm}
F.~Denef, M.~R. Douglas, B.~Florea, A.~Grassi, and S.~Kachru, {\it {Fixing All
  Moduli in a Simple F-Theory Compactification}},  {\em Adv. Theor. Math.
  Phys.} {\bf 9} (2005) 861--929,
  [\href{http://xxx.lanl.gov/abs/hep-th/0503124}{{\tt hep-th/0503124}}].

\bibitem{BBCQ}
V.~Balasubramanian, P.~Berglund, J.~P. Conlon, and F.~Quevedo, {\it
  {Systematics of Moduli Stabilisation in Calabi-Yau Flux Compactifications}},
  {\em JHEP} {\bf 03} (2005) 007,
  [\href{http://xxx.lanl.gov/abs/hep-th/0502058}{{\tt hep-th/0502058}}].

\bibitem{BHP}
M.~Berg, M.~Haack, and E.~Pajer, {\it {Jumping Through Loops: On Soft Terms
  from Large Volume Compactifications}},  {\em JHEP} {\bf 09} (2007) 031,
  [\href{http://xxx.lanl.gov/abs/0704.0737}{{\tt arXiv:0704.0737}}].

\bibitem{CQS}
J.~P. Conlon, F.~Quevedo, and K.~Suruliz, {\it {Large-Volume Flux
  Compactifications: Moduli Spectrum and D3/D7 Soft Supersymmetry Breaking}},
  {\em JHEP} {\bf 08} (2005) 007,
  [\href{http://xxx.lanl.gov/abs/hep-th/0505076}{{\tt hep-th/0505076}}].

\bibitem{KKLMMT}
S.~Kachru, R.~Kallosh, A.~Linde, J.~Maldacena, L.~McAllister, and S.~P.
  Trivedi, {\it {Towards Inflation in String Theory}},  {\em JCAP} {\bf 0310}
  (2003) 013, [\href{http://xxx.lanl.gov/abs/hep-th/0308055}{{\tt
  hep-th/0308055}}].

\bibitem{BlancoPillado:2004ns}
J.~J. Blanco-Pillado {\em et.~al.}, {\it {Racetrack Inflation}},  {\em JHEP}
  {\bf 11} (2004) 063, [\href{http://xxx.lanl.gov/abs/hep-th/0406230}{{\tt
  hep-th/0406230}}].

\bibitem{BDKM}
D.~Baumann, A.~Dymarsky, I.~R. Klebanov, and L.~McAllister, {\it {Towards an
  Explicit Model of D-brane Inflation}},  {\em JCAP} {\bf 0801} (2008) 024,
  [\href{http://xxx.lanl.gov/abs/0706.0360}{{\tt arXiv:0706.0360}}].

\bibitem{CQ-infl}
J.~P. Conlon and F.~Quevedo, {\it {K\"ahler Moduli Inflation}},  {\em JHEP} {\bf
  01} (2006) 146, [\href{http://xxx.lanl.gov/abs/hep-th/0509012}{{\tt
  hep-th/0509012}}].

\bibitem{KP}
A.~Krause and E.~Pajer, {\it {Chasing Brane Inflation in String-Theory}},
  [\href{http://xxx.lanl.gov/abs/0705.4682}{{\tt arXiv:0705.4682}}].

\bibitem{D&DI}
F.~Denef and M.~R. Douglas, {\it Distributions of Flux Vacua},  {\em JHEP} {\bf
  05} (2004) 072, [\href{http://xxx.lanl.gov/abs/hep-th/0404116}{{\tt
  hep-th/0404116}}].

\bibitem{BB}
V.~Balasubramanian and P.~Berglund, {\it {Stringy Corrections to K\"ahler
  Potentials, SUSY Breaking, and the Cosmological Constant Problem}},  {\em
  JHEP} {\bf 11} (2004) 085,
  [\href{http://xxx.lanl.gov/abs/hep-th/0408054}{{\tt hep-th/0408054}}].

\bibitem{SS}
A.~Saltman and E.~Silverstein, {\it {The Scaling of the No-Scale Potential and
  de Sitter Model Building}},  {\em JHEP} {\bf 11} (2004) 066,
  [\href{http://xxx.lanl.gov/abs/hep-th/0402135}{{\tt hep-th/0402135}}].

\bibitem{DG}
O.~DeWolfe and S.~B. Giddings, {\it {Scales and Hierarchies in Warped
  Compactifications and Brane Worlds}},  {\em Phys. Rev.} {\bf D67} (2003)
  066008, [\href{http://xxx.lanl.gov/abs/hep-th/0208123}{{\tt
  hep-th/0208123}}].

\bibitem{Grana:2003ek}
M.~Grana, T.~W. Grimm, H.~Jockers, and J.~Louis, {\it {Soft Supersymmetry
  Breaking in Calabi-Yau Orientifolds with D-Branes and Fluxes}},  {\em Nucl.
  Phys.} {\bf B690} (2004) 21--61,
  [\href{http://xxx.lanl.gov/abs/hep-th/0312232}{{\tt hep-th/0312232}}].

\bibitem{Baumann:2006th}
D.~Baumann, A.~Dymarsky, I.~R. Klebanov, J.~M. Maldacena, L.~P. McAllister, and
  A.~Murugan, {\it {On D3-Brane Potentials in Compactifications with Fluxes and
  Wrapped D-Branes}},  {\em JHEP} {\bf 11} (2006) 031,
  [\href{http://xxx.lanl.gov/abs/hep-th/0607050}{{\tt hep-th/0607050}}].

\bibitem{GVW}
S.~Gukov, C.~Vafa, and E.~Witten, {\it {CFT's from Calabi-Yau Four-Folds}},
  {\em Nucl. Phys.} {\bf B584} (2000) 69--108,
  [\href{http://xxx.lanl.gov/abs/hep-th/9906070}{{\tt hep-th/9906070}}].

\bibitem{Lust:2005bd}
D.~Lust, P.~Mayr, S.~Reffert, and S.~Stieberger, {\it {F-Theory Flux,
  Destabilization of Orientifolds and Soft Terms on D7-branes}},  {\em Nucl.
  Phys.} {\bf B732} (2006) 243--290,
  [\href{http://xxx.lanl.gov/abs/hep-th/0501139}{{\tt hep-th/0501139}}].

\bibitem{CDE}
A.~Collinucci, F.~Denef, and M.~Esole, {\it {D-brane Deconstructions in IIB
  Orientifolds}},  [\href{http://xxx.lanl.gov/abs/0805.1573}{{\tt
  arXiv:0805.1573}}].

\bibitem{Grana:2005jc}
M.~Grana, {\it {Flux compactifications in String Theory: A Comprehensive
  Review}},  {\em Phys. Rept.} {\bf 423} (2006) 91--158,
  [\href{http://xxx.lanl.gov/abs/hep-th/0509003}{{\tt hep-th/0509003}}].

\bibitem{Douglas:2006es}
M.~R. Douglas and S.~Kachru, {\it {Flux Compactification}},  {\em Rev. Mod.
  Phys.} {\bf 79} (2007) 733--796,
  [\href{http://xxx.lanl.gov/abs/hep-th/0610102}{{\tt hep-th/0610102}}].

\bibitem{Berg:2005ja}
M.~Berg, M.~Haack, and B.~Kors, {\it {String Loop Corrections to K\"ahler
  Potentials in Orientifolds}},  {\em JHEP} {\bf 11} (2005) 030,
  [\href{http://xxx.lanl.gov/abs/hep-th/0508043}{{\tt hep-th/0508043}}].

\bibitem{Sen-FO}
A.~Sen, {\it {F-Theory and Orientifolds}},  {\em Nucl. Phys.} {\bf B475} (1996)
  562--578, [\href{http://xxx.lanl.gov/abs/hep-th/9605150}{{\tt
  hep-th/9605150}}].

\bibitem{GL}
T.~W. Grimm and J.~Louis, {\it {The Effective Action of N = 1 Calabi-Yau
  Orientifolds}},  {\em Nucl. Phys.} {\bf B699} (2004) 387--426,
  [\href{http://xxx.lanl.gov/abs/hep-th/0403067}{{\tt hep-th/0403067}}].

\bibitem{Str-SG}
A.~Strominger, {\it {Special Geometry}},  {\em Commun. Math. Phys.} {\bf 133}
  (1990) 163--180.

\bibitem{Candelas:1990pi}
P.~Candelas and X.~de~la Ossa, {\it {Moduli Space of Calabi-Yau Manifolds}},
  {\em Nucl. Phys.} {\bf B355} (1991) 455--481.

\bibitem{Becker:2007ee}
K.~Becker, Y.-C. Chung, and G.~Guo, {\it {Metastable Flux Configurations and de
  Sitter Spaces}},  {\em Nucl. Phys.} {\bf B790} (2008) 240--257,
  [\href{http://xxx.lanl.gov/abs/0706.2502}{{\tt arXiv:0706.2502}}].

\bibitem{Kallosh:2005ax}
R.~Kallosh, {\it {New Attractors}},  {\em JHEP} {\bf 12} (2005) 022,
  [\href{http://xxx.lanl.gov/abs/hep-th/0510024}{{\tt hep-th/0510024}}].

\bibitem{Bellucci:2007ds}
S.~Bellucci, S.~Ferrara, R.~Kallosh, and A.~Marrani, {\it {Extremal Black Hole
  and Flux Vacua Attractors}},  [\href{http://xxx.lanl.gov/abs/0711.4547}{{\tt
  arXiv:0711.4547}}].

\bibitem{D&DII}
F.~Denef and M.~R. Douglas, {\it {Distributions of Nonsupersymmetric Flux
  Vacua}},  {\em JHEP} {\bf 03} (2005) 061,
  [\href{http://xxx.lanl.gov/abs/hep-th/0411183}{{\tt hep-th/0411183}}].

\bibitem{WB}
J.~Wess and J.~Bagger, {\it {Supersymmetry and Supergravity}}, . Princeton,
  USA: Univ. Pr. (1992) 259 p.

\bibitem{D0}
M.~R. Douglas, {\it {The Statistics of String / M Theory Vacua}},  {\em JHEP}
  {\bf 05} (2003) 046, [\href{http://xxx.lanl.gov/abs/hep-th/0303194}{{\tt
  hep-th/0303194}}].

\bibitem{AD}
S.~Ashok and M.~R. Douglas, {\it {Counting Flux Vacua}},  {\em JHEP} {\bf 01}
  (2004) 060, [\href{http://xxx.lanl.gov/abs/hep-th/0307049}{{\tt
  hep-th/0307049}}].

\bibitem{Candelas:1990rm}
P.~Candelas, X.~C. De~La~Ossa, P.~S. Green, and L.~Parkes, {\it {A Pair of
  Calabi-Yau Manifolds as an Exactly Soluble Superconformal Theory}},  {\em
  Nucl. Phys.} {\bf B359} (1991) 21--74.

\bibitem{Hosono:1994av}
S.~Hosono, A.~Klemm, and S.~Theisen, {\it {Lectures on Mirror Symmetry}},
  [\href{http://xxx.lanl.gov/abs/hep-th/9403096}{{\tt hep-th/9403096}}].

\bibitem{C-MSI}
P.~Candelas, X.~De~La~Ossa, A.~Font, S.~H. Katz, and D.~R. Morrison, {\it
  Mirror Symmetry for Two Parameter Models, I},  {\em Nucl. Phys.} {\bf B416}
  (1994) 481--538, [\href{http://xxx.lanl.gov/abs/hep-th/9308083}{{\tt
  hep-th/9308083}}].

\bibitem{Misra:2004ky}
A.~Misra and A.~Nanda, {\it {Flux Vacua Statistics for Two-Parameter
  Calabi-Yau's}},  {\em Fortsch. Phys.} {\bf 53} (2005) 246--259,
  [\href{http://xxx.lanl.gov/abs/hep-th/0407252}{{\tt hep-th/0407252}}].

\bibitem{Kaura:2006mv}
P.~Kaura and A.~Misra, {\it {On the Existence of Non-Supersymmetric Black Hole
  Attractors for Two-Parameter Calabi-Yau's and Attractor Equations}},  {\em
  Fortsch. Phys.} {\bf 54} (2006) 1109--1141,
  [\href{http://xxx.lanl.gov/abs/hep-th/0607132}{{\tt hep-th/0607132}}].

\bibitem{Conlon:2004ds}
J.~P. Conlon and F.~Quevedo, {\it {On the Explicit Construction and Statistics
  of Calabi-Yau Flux Vacua}},  {\em JHEP} {\bf 10} (2004) 039,
  [\href{http://xxx.lanl.gov/abs/hep-th/0409215}{{\tt hep-th/0409215}}].

\bibitem{CCQ}
M.~Cicoli, J.~P. Conlon, and F.~Quevedo, {\it {Systematics of String Loop
  Corrections in Type IIB Calabi- Yau Flux Compactifications}},  {\em JHEP}
  {\bf 01} (2008) 052, [\href{http://xxx.lanl.gov/abs/0708.1873}{{\tt
  arXiv:0708.1873}}].

\bibitem{Becker:2002nn}
K.~Becker, M.~Becker, M.~Haack, and J.~Louis, {\it {Supersymmetry Breaking and
  $\alpha'$-corrections to Flux Induced Potentials}},  {\em JHEP} {\bf 06} (2002)
  060, [\href{http://xxx.lanl.gov/abs/hep-th/0204254}{{\tt hep-th/0204254}}].

\bibitem{Berg:2005yu}
M.~Berg, M.~Haack, and B.~Kors, {\it {On Volume Stabilization by Quantum
  Corrections}},  {\em Phys. Rev. Lett.} {\bf 96} (2006) 021601,
  [\href{http://xxx.lanl.gov/abs/hep-th/0508171}{{\tt hep-th/0508171}}].

\bibitem{Burgess:2005jx}
C.~P. Burgess, C.~Escoda, and F.~Quevedo, {\it {Nonrenormalization of Flux
  Superpotentials in String Theory}},  {\em JHEP} {\bf 06} (2006) 044,
  [\href{http://xxx.lanl.gov/abs/hep-th/0510213}{{\tt hep-th/0510213}}].

\bibitem{Berglund:2005dm}
P.~Berglund and P.~Mayr, {\it {Non-Perturbative Superpotentials in F-Theory and
  String Duality}},  [\href{http://xxx.lanl.gov/abs/hep-th/0504058}{{\tt
  hep-th/0504058}}].

\bibitem{JL}
H.~Jockers and J.~Louis, {\it {The Effective Action of D7-branes in N = 1
  Calabi-Yau Orientifolds}},  {\em Nucl. Phys.} {\bf B705} (2005) 167--211,
  [\href{http://xxx.lanl.gov/abs/hep-th/0409098}{{\tt hep-th/0409098}}].

\bibitem{Gorlich:2004qm}
L.~Gorlich, S.~Kachru, P.~K. Tripathy, and S.~P. Trivedi, {\it {Gaugino
  Condensation and Nonperturbative Superpotentials in Flux Compactifications}},
   {\em JHEP} {\bf 12} (2004) 074,
  [\href{http://xxx.lanl.gov/abs/hep-th/0407130}{{\tt hep-th/0407130}}].

\bibitem{deAlwis:2005tf}
S.~P. de~Alwis, {\it {Effective Potentials for Light Moduli}},  {\em Phys.
  Lett.} {\bf B626} (2005) 223--229,
  [\href{http://xxx.lanl.gov/abs/hep-th/0506266}{{\tt hep-th/0506266}}].

\bibitem{Misra:2007yu}
A.~Misra and P.~Shukla, {\it {Area Codes, Large Volume (Non-)Perturbative
  alpha'- and Instanton - Corrected Non-supersymmetric (A)dS minimum, the
  Inverse Problem and Fake Superpotentials for Multiple-
  Singular-Loci-Two-Parameter Calabi-Yau's}},
  [\href{http://xxx.lanl.gov/abs/0707.0105}{{\tt arXiv:0707.0105}}].

\bibitem{CGGLPS}
L.~Covi, M.~Gomez-Reino, C.~Gross, J.~Louis, G.~A. Palma, and C.~A. Scrucca,
  {\it {De Sitter vacua in No-Scale Supergravities and Calabi-Yau String
  Models}},  {\em JHEP} {\bf 06} (2008) 057,
  [\href{http://xxx.lanl.gov/abs/0804.1073}{{\tt arXiv:0804.1073}}].

\bibitem{HMR}
A.~Hebecker and J.~March-Russell, {\it {The Ubiquitous Throat}},  {\em Nucl.
  Phys.} {\bf B781} (2007) 99--111,
  [\href{http://xxx.lanl.gov/abs/hep-th/0607120}{{\tt hep-th/0607120}}].

\bibitem{Soroush}
M.~Soroush, {\it {Constraints on meta-stable de Sitter flux vacua}},
  [\href{http://xxx.lanl.gov/abs/hep-th/0702204}{{\tt hep-th/0702204}}].

\bibitem{BF}
P.~Breitenlohner and D.~Z. Freedman, {\it {Positive Energy in Anti-de Sitter
  Backgrounds and Gauged Extended Supergravity}},  {\em Phys. Lett.} {\bf B115}
  (1982) 197.

\bibitem{Chialva:2007sv}
D.~Chialva, U.~H. Danielsson, N.~Johansson, M.~Larfors, and M.~Vonk, {\it
  {Deforming, Revolving and Resolving - New Paths in the String Theory
  Landscape}},  {\em JHEP} {\bf 02} (2008) 016,
  [\href{http://xxx.lanl.gov/abs/0710.0620}{{\tt 0710.0620}}].

\bibitem{Johnson:2008kc}
M.~C. Johnson and M.~Larfors, {\it {Field Dynamics and Tunneling in a Flux
  Landscape}},  [\href{http://xxx.lanl.gov/abs/0805.3705}{{\tt
  arXiv:0805.3705}}].

\end{thebibliography}\endgroup

\end{document}